\documentclass[english]{iopart}
\usepackage[pdftex]{graphicx}
\usepackage{color}
\usepackage{esint}
\usepackage{iopams,amssymb,bm}
\usepackage{setstack}
\usepackage{braket}
\usepackage{babel}
\def\vec#1{{\bf#1}}
\def\uvec#1{\hat{\bf#1}}
\def\gvec#1{{\bm#1}}

\def\op#1{#1}
\def\ip#1#2{\langle #1 \mid #2 \rangle}

\def\norm#1{\| #1 \|}

\def\H{\mathcal{H}}
\def\RR{\mathbb{R}}

\def\CC{\mathbb{C}}
\def\ONE{\mathbb{I}}

\def\F{\mathcal{F}}
\def\eps{\epsilon}
\def\sx{\op{\sigma}_x}
\def\sy{\op{\sigma}_y}
\def\sz{\op{\sigma}_z}

\def\Had{\mbox{\rm Had}}

\begin{document}
\title{Implementation of Fault Tolerant Quantum Logic Gates via Optimal Control}
\author{R Nigmatullin$^{a},$ S Schirmer$^{b,c}$}
\address{$^{a}$Cavendish Laboratory, University of Cambridge, J J Thomson
Avenue, Cambridge, CB3 0HE, United Kingdom}
\address{$^{b}$Dept of Applied Maths and Theoretical Physics, University
of Cambridge, Wilberforce Road, CB3 0WA, United Kingdom}
\address{$^{c}$Kavli Institute of Theoretical Physics, UCSB, Santa
Barbara, CA93106, USA}
\ead{sgs29@cam.ac.uk}

\begin{abstract}
The implementation of fault-tolerant quantum gates on encoded logic
qubits is considered.  It is shown that transversal implementation of
logic gates based on simple geometric control ideas is problematic for
realistic physical systems suffering from imperfections such as qubit
inhomogeneity or uncontrollable interactions between qubits.  However,
this problem can be overcome by formulating the task as an optimal
control problem and designing efficient algorithms to solve it.  In
particular, we can find solutions that implement all of the elementary
logic gates in a fixed amount of time with limited control resources for
the five-qubit stabilizer code.  Most importantly, logic gates that are
extremely difficult to implement using conventional techniques even for
ideal systems, such as the $T$-gate for the five-qubit stabilizer code,
do not appear to pose a problem for optimal control.
\end{abstract}
\maketitle

\section{Introduction}

Quantum information processing has been a topic of intense theoretical
and experimental research for several years.  Many potential physical
realizations of quantum computers have been proposed, and although the
experimental realization of quantum information processing remains a
challenge, there have been many experimental accomplishments including
the demonstration of control of multi-qubit dynamics in spin systems
using nuclear magnetic resonance techniques~\cite{NMR2004} and
all-optical quantum information processing using photons~\cite{BRI2008},
for instance.  While liquid-state NMR and optical quantum computing may
suffer from inherent scalability issues, there has also been significant
progress in ion-trap architectures and several proposals for making such
architectures scalable exist~\cite{WIN2007, WIN2008}.  In solid-state
systems successes have been more modest but controlled interactions of
quantum dots have been demonstrated in some
systems~\cite{SMITH2007,IMA2008}.

One major obstacle to scalability is the increasing difficulty in
effectively controlling the dynamics of many qubits as the complexity of
experimental systems increases.  While it is in principle easy to
control a single two-level system, the situation is more complicated
when many qubits are involved.  In many cases individual addressability
of qubits in a register or array is difficult to achieve; inhomogeneity
may result in different qubits having different responses to control
fields, and the dynamics is complicated by uncontrollable inter-qubit
couplings.  All of these issues present challenges, even in the absence
of environmental noise and decoherence.  These problems are magnified
because the necessity to be able to correct at least a certain amount of
inevitable errors means that bits of quantum information must be encoded
in logical qubits consisting of multiple physical qubits to ensure that
there is sufficient redundancy.  Even if only the most basic level of
protection is assumed, at least five physical qubits are required to
encode a single logical qubit so that we can correct a single (bit or
phase flip) error~\cite{NIE2000}.  In practice multiple layers of error
correction will be necessary to achieve fault-tolerant operation of a
quantum processor for a sufficiently long time to allow the completion
of a non-trivial quantum algorithm~\cite{NIE2000}.  Thus, in the setting
of fault-tolerant computation using encoded qubits even the
implementation of single qubit logic gates becomes a non-trivial
multi-qubit control problem.

The implementation of fault-tolerant gates quantum logic gates seems to
be a problem that is well suited for optimal control.  Optimal control
theory has been used in various papers to implement two and three qubit
gates as well as simple quantum circuits, and has been shown to improve
gate fidelities, gate operation times and robustness
(e.g.~\cite{KHAN2001,HERB2005,SCHI2009}).  In this paper we consider the
implementation of logic gates on encoded qubits, which is a next logical
step.  Although most logic gates on encoded qubits for the most common
codes can be implemented, in principle, by applying a particular single
qubit operation to all physical qubits, what is known as transversal
implementation, not all logic gates can be implemented this way even in
the ideal case, and transversal implementation is problematic in the
presence of inhomogeneity or uncontrollable couplings between physical
qubits, and the implementation of encoded logic gates in this setting is
therefore in an interesting challenge for optimal control.  In
particular we study the implementation of a set of logic gates for the
five-qubit stabilizer code for model systems with both inhomogeneity and
fixed inter-qubit couplings.  The paper is organized as follows.  In
Sec.~\ref{sec:encoding} we briefly introduce stabilizer codes and the
construction of fault-tolerant logic gates with emphasis on the
five-qubit stabilizer code that will be the main focus of this paper.
In Sec.~\ref{sec:models} we describe two types of model systems and
discuss the implementation of encoded logic gates for these system based
on geometric control ideas.  In section~\ref{sec:OCT} we formulate the
optimal control problem and show how to design iterative algorithms to
solve it numerically.  In Sec.~\ref{sec:applic} the results are applied
to find optimal controls to implement a set of standard logic gates for
the five-qubit stabilizer code for two classes of model systems, and the
merits and drawback of the optimal control approach compared with the
simpler geometric control schemes are discussed.

\section{Stabilizer codes and fault-tolerant gates}
\label{sec:encoding}

The state of a single qubit can be represented by a density operator
$\rho$ on a Hilbert space $\H\simeq \CC^2$, which can be expanded with
respect to the standard Pauli basis,
$\rho=\frac{1}{2}(\ONE+x\sx+y\sy+z\sz)$, where $\ONE$ is the identity
matrix and
\begin{equation}
  \sx= \left(\begin{array}{cc} 0 & 1 \\ 1 & 0 \end{array}\right),\quad
  \sy= \left(\begin{array}{cc} 0 & -i\\ i & 0 \end{array}\right),\quad
  \sz= \left(\begin{array}{cc} 1 & 0 \\ 0 & -1\end{array}\right).
\end{equation}
are the usual Pauli matrices, and we have $*=\Tr(\sigma_*\rho)$ for
$*\in\{x,y,z\}$.  Any single qubit gate can be written
\begin{equation} 
  R_{\uvec{n}}(\alpha)
  =\exp\left(i\frac{\alpha}{2}\, \uvec{n}\cdot\gvec{\sigma}\right),
  \label{eq:Grotation}
\end{equation}
where $\gvec{\sigma}=(\sx,\sy,\sz)$, which can be interpreted as a
rotation of the Bloch vector $\vec{s}=(x,y,z)\in\RR^3$ by an angle
$\alpha$ around the axis $\uvec{n}=(n_x,n_y,n_z)$.  We further note that
the norm of the Bloch vector $\norm{\vec{s}}\le 1$, with equality if and
only if $\rho$ represents a pure state.

The ability to implement universal quantum operations on $n$ qubits
requires a minimal set of elementary gates, which typically comprises
several essential single qubit gates such as the identity $\ONE$, the 
$S$ and $T$ gates and the Hadamard gate
\begin{equation}
  S=R_{\uvec{z}}\left(\frac{\pi}{2}\right), \quad 
  T=R_{\uvec{z}}\left(\frac{\pi}{4}\right), \quad
  \Had =  R_{\uvec{n}}(\pi),
\end{equation}
where $\uvec{n}=\frac{1}{\sqrt{2}}(1,0,1)$, and a single universal
two-qubit gate such as the CNOT gate~\cite{NIE2000}.  Often the
$\pi$-rotations about the $x$, $y$ and $z$-axis, $X=\sx$, $Y=\sy$,
$Z=\sz$, respectively, are also included for convenience.

\begin{figure}
\center
\includegraphics[width=\columnwidth]{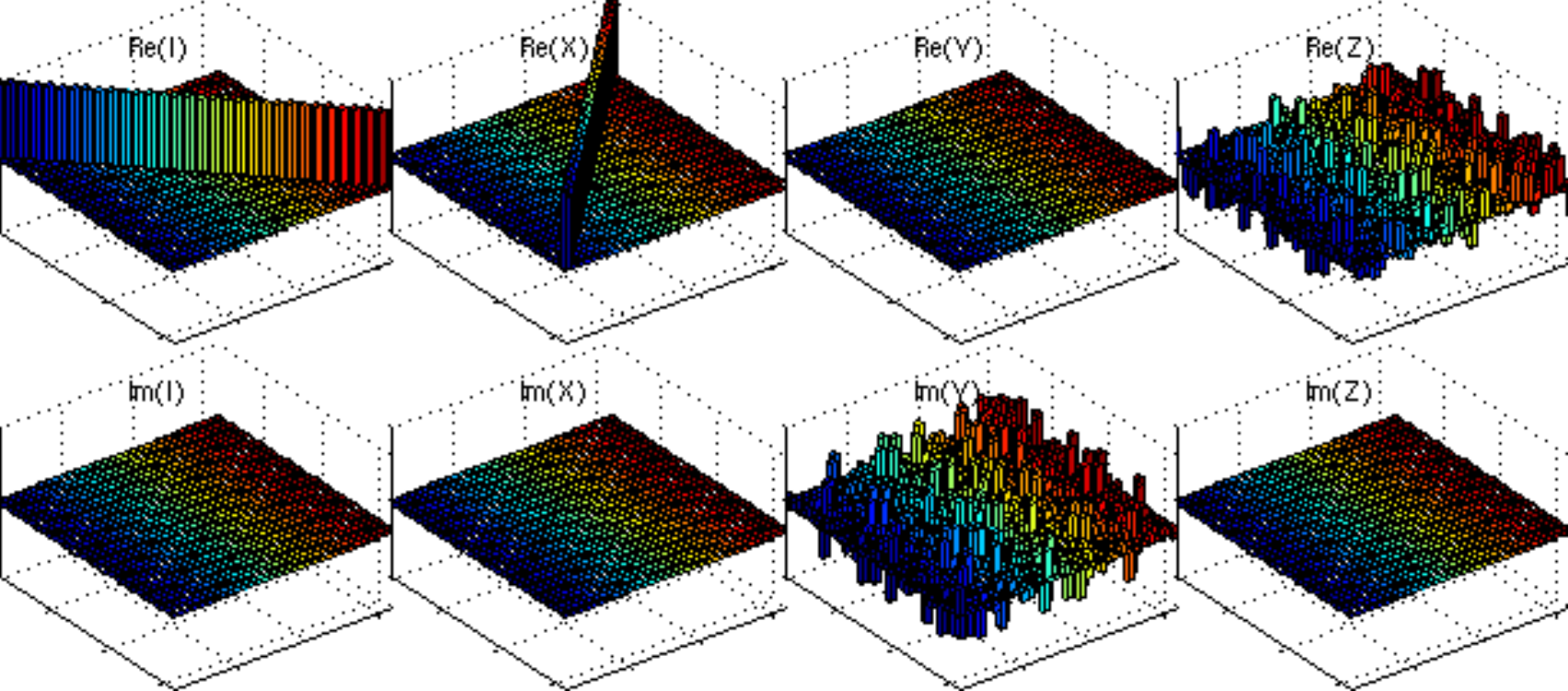}
\includegraphics[width=\columnwidth]{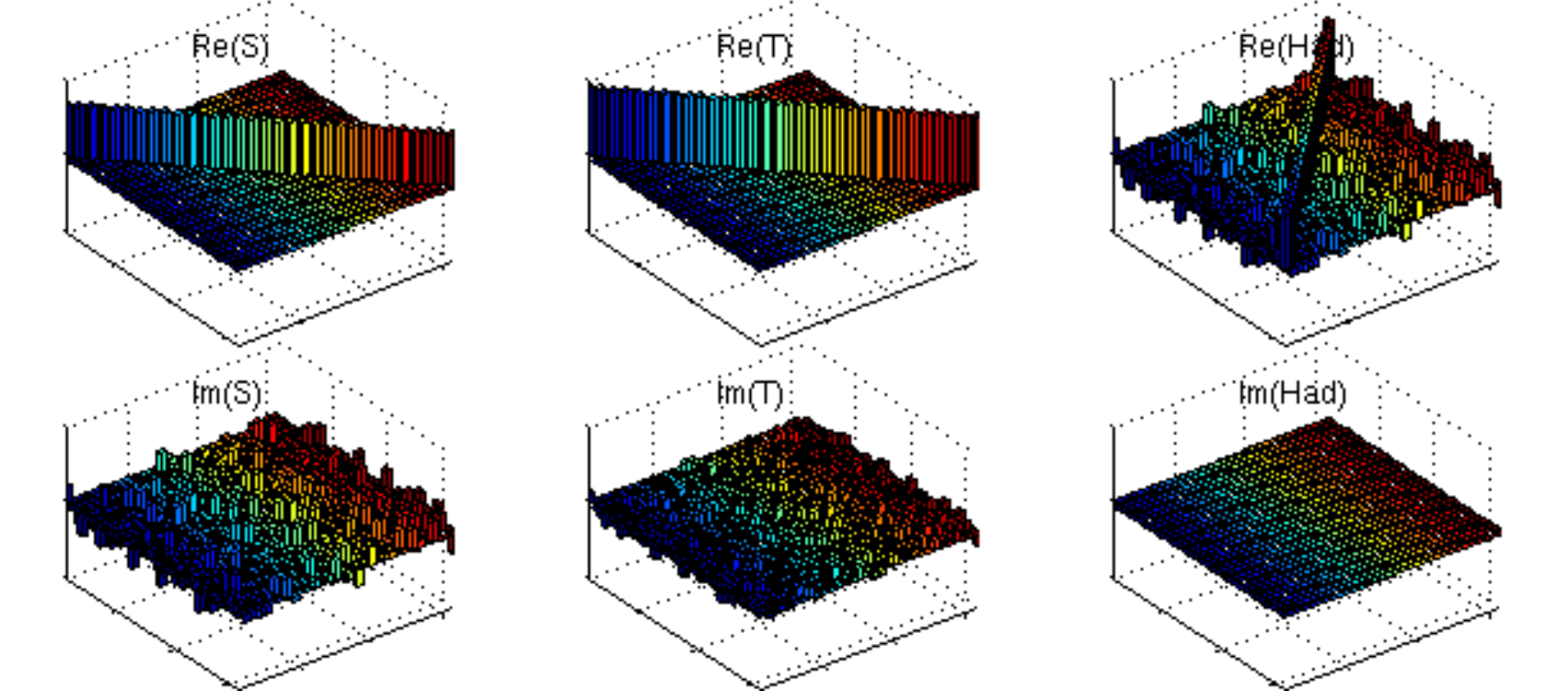}
\caption{Real and imaginary parts of $32\times 32$ unitary operators
corresponding to the encoded logic gates.}
\label{fig:target_gates}
\end{figure}

To protect quantum information from errors due to noise and decoherence
several physical qubits are required to encode a logical qubit in a way
that enables us to recover quantum information from corrupted qubits.
Different types of encodings exists to protect against different types
of errors, but the most common encodings are decoherence-free subspaces
and stabilizer codes~\cite{NIE2000}.  In this paper we focus on the
latter type of encoding, where errors can be corrected by performing
(projective) syndrome measurements on encoded qubits and applying the
correcting quantum gates to the corrupted qubit.  The minimum number of
physical qubits required to encode a single quantum bit of information
such that we can recover from a single bit or phase flip error on any of
the physical qubits is the five-qubit stabilizer code based on the encoding
\begin{eqnarray}
\ket{0_{L}} 
 &=& \frac{1}{4}[\ket{00000}+\ket{10010}+\ket{01001}+\ket{10100}\nonumber \\
 & & +\ket{01010}-\ket{11011}-\ket{00110}-\ket{11000}\nonumber \\
 & & -\ket{11101}-\ket{00011}-\ket{11110}-\ket{01111}\nonumber \\
 & & -\ket{10001}-\ket{01100}-\ket{10111}-\ket{00101}] \\
\ket{1_{L}} 
 &=& \frac{1}{4}[\ket{11111}+\ket{01101}+\ket{10110}+\ket{01011}\nonumber \\
 & & +\ket{10101}-\ket{00100}-\ket{11001}-\ket{00111}\nonumber \\
 & & -\ket{00010}-\ket{11100}-\ket{00001}-\ket{10000}\nonumber \\
 & & -\ket{01110}-\ket{10011}-\ket{01000}-\ket{11010}],
\end{eqnarray}
where $\ket{00\ldots}$ is the usual shorthand for the tensor product
$\ket{0}\otimes\ket{0}\otimes\ldots$.  Thus, a single qubit logic gate
on encoded qubits is equivalent to a five-qubit gate represented by a
$32\times 32$ unitary matrix.  There is a certain degree of freedom in
the definition of logic gates but a logic $X$-gate, for instance, must
obviously swap the logic states $\ket{0_L}$ and $\ket{1_L}$.  Moreover,
the key premise of fault-tolerant logic gate construction is that the
application of a particular logic gate to a corrupted qubit, should not
increase the number of errors.  Thus, an encoded qubit corrupted by a
single bit or phase flip error on a physical qubit must be mapped to a
state with a single bit or phase flip error to enable subsequent error
correction.  If $X_n$ is denotes a bit flip applied to the $n$th qubit
then the requirement of unitarity prevents us from directly correcting
errors as we cannot map both $\ket{0_L}$ and the corrupted state
$X_n\ket{0_L}$ to $\ket{1_L}$.  With this in mind, it is straightforward
to derive suitable matrix representations for the fault-tolerant gates.
For example, the logical $X_L$ gate must map $\ket{0_{L}}$ to
$\ket{1_{L}}$ and vice versa.  Furthermore, an input state corrupted by
a single bit flip error on the $n$th qubit, $X_{n}\ket{0_{L}}$ should be
mapped to the output state $X_{n}\ket{1_{L}}$, i.e., the target state
$\ket{1_L}$ with a bit flip on the $n$th qubit, or more generally
\begin{equation} \fl
X_{L} = \ket{1_{L}}\bra{0_{L}}
        +\sum_{n>m}[X_{n}\ket{1_{L}}\bra{0_{L}}X_{n}
                  +X_{n}X_{m}\ket{1_{L}}\bra{0_{L}}X_{m}X_{n}]+c.c.
\label{eq:logicalX}
\end{equation}
where $c.c.$ denotes the complex conjugate.  The same relationship
between the input and output states should hold when the bit flip
errors $X_n$ are replaced by phase flip errors $Z_n$.  This fully
determines the logic $X_L$ gate.  We can similarly derive explicit
representations for all other logic gates.  Fig.~\ref{fig:target_gates}
shows the structure of the resulting unitary operators corresponding to
the target gates.

Given a register of identical and non-interacting qubits, most
fault-tolerant quantum logic gates can be implemented transversally,
i.e., by applying the same single qubit operation in parallel to each
qubit in the register.  For instance, it is can easily be verified that
the logical $X$-gate defined by~(\ref{eq:logicalX}) is simply $X_L =
X\otimes X\otimes X\otimes X\otimes X$, i.e., an $X$ gate applied to
each physical qubit, However, even if we are given ideal non-interacting
physical qubits, not all quantum logic gates can be implemented
transversally.  For instance, the $T$-gate above, which is required for
universality, cannot be realized this way.  Indeed, implementation of
the $T$-gate for the five-qubit code above is generally complicated,
requiring auxiliary qubits and quantum teleportation~\cite{NIE2000}.
Furthermore, for real physical systems inhomogeneity and uncontrollable
interactions may make simple transverse implementation of fault-tolerant
gates impractical, if not impossible.

\section{Model systems and geometric control schemes}
\label{sec:models}

Quite a few proposed realizations for quantum computing architectures
can be formally thought of as linear arrays of two-level systems (qubits
or pseudo-spin-$\frac{1}{2}$-particles).  In the ideal case one usually
considers individual spins that are identical, individually addressable
and controllable, and assumes the couplings between all qubits are fully
controllable.  In practice, however, these requirements are difficult to
meet as the coupling between spins is often not controllable, and the
qubits may not be identical or selectively addressable.  In this case
simple transverse implementation of quantum logic gates is not possible
even for simple gates such as the $X$-gate, and an optimal control
approach seems the most promising way to overcome such obstacles to
implement quantum logic gates with limited control.

\begin{figure}
\center\includegraphics[width=0.8\columnwidth]{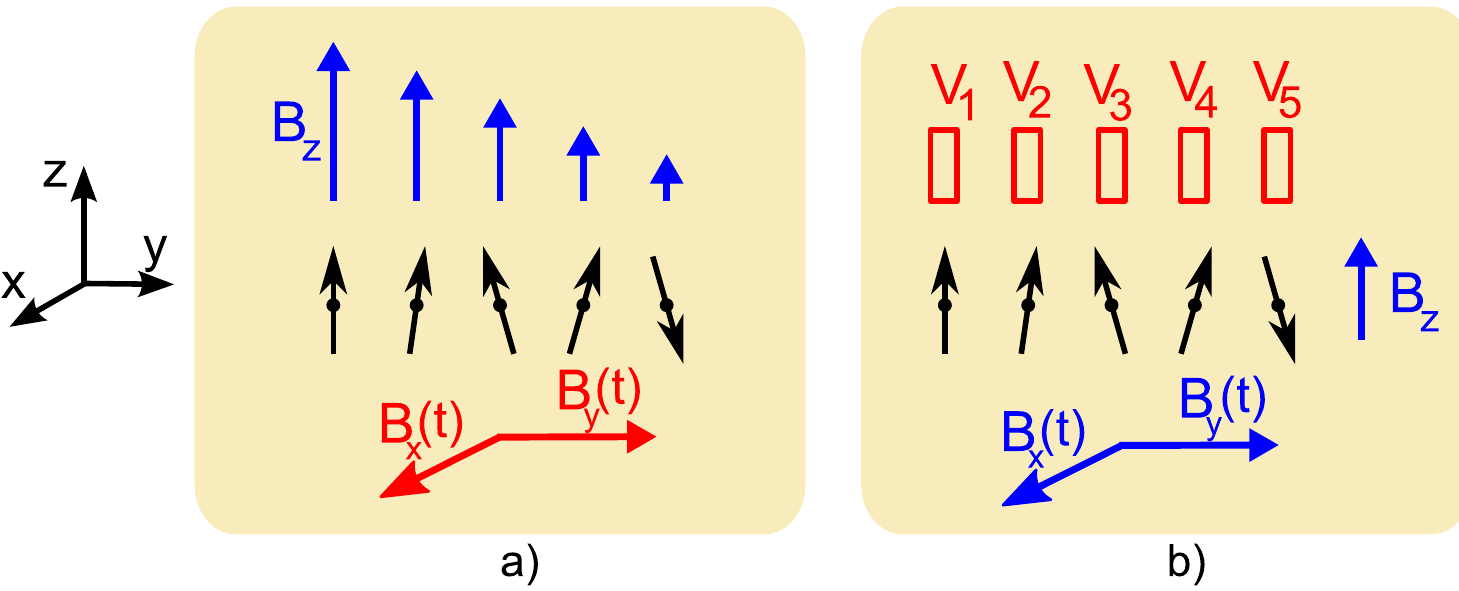}
\caption{Spin chain controlled (a) by modulation of global fields and
(b) using a fixed global field and local electrodes that tune individual
qubits in and out of resonance with the field.  $B_x(t)$, $B_y(t)$ and
$B_z(t)$ indicate the $x$, $y$ and $z$ components of a globally applied
field.  $V_k$ refer to control parameters for local actuators, e.g.,
voltages applied to control electrodes.} \label{fig:Ising-chain}
\end{figure}

In the following we consider two control paradigms (a) global control
and (b) limited local control, as shown in Fig.~\ref{fig:Ising-chain}.
In case (a) the physical qubits comprising the logic qubit are
controlled entirely by a global field that interacts with all physical
qubits simultaneously.  In case (b) there may also be a fixed global
field but the actual control operations are performed using local
control gates, such as local control electrodes that tune individual
spins in and out of resonance with a global field.

\subsection{Global Control}

If the qubits are identical and there are no couplings between them,
then transversal logic gates can be easily implemented in the global
control setting by simply finding a control pulse that implements the
desired gate for a single qubit and applying it to all physical qubits.
The elementary single qubit gates can be implemented, e.g., by applying
a sequence of simple pulses effecting rotations about two orthogonal
axes, using the Euler decomposition
\begin{equation}
  R_{\uvec{n}}(\theta)
  = R_{\uvec{x}}(\alpha) R_{\uvec{y}}(\beta) R_{\uvec{x}}(\gamma)
\end{equation}
for suitably chosen values of $\alpha,$ $\beta$ and $\gamma$, as shown
in Table~\ref{table1}.  Rotations about the $x$- or $y$-axis can be
performed by applying a pulse of the form $f(t)=B(t)\cos(\omega t+\phi)$
with suitable frequency $\omega$, pulse area $\int_{t_0}^{t_F} B(t)\,dt$ 
and phase $\phi$~\cite{SCHI2009}.

\begin{table}
\begin{center}
\begin{tabular}{|c|c|c|}
\hline 
Gate & Euler decomposition & Pulse length \tabularnewline
\hline\hline 
X & $R_{\uvec{x}}(\pi)$                            & 1\\\hline 
Y & $R_{\uvec{y}}(\pi)$                            & 1\\\hline 
Z & $R_{\uvec{x}}(\pi)R_{\uvec{y}}(\pi)$            & 2\\\hline 
I & $R_{\uvec{x}}(2\pi)$                           & 2\\\hline 
S & $R_{\uvec{x}}(3\pi/2)R_{\uvec{y}}(\pi/2)R_{\uvec{x}}(\pi/2)$ & 2.50\\\hline 
T & $R_{\uvec{x}}(3\pi/2)R_{\uvec{y}}(\pi/4)R_{\uvec{x}}(\pi/2)$ & 2.25\\\hline 
$\Had$ & $R_{\uvec{y}}(\pi/2)R_{\uvec{x}}(\pi)$       & 1.50\\\hline
\end{tabular}
\end{center}
\caption{Euler decomposition of elementary gates in terms of rotations
about $\uvec{x}$ and $\uvec{y}$ axis and total pulse length in units of
$2\Omega/\pi$, where $\Omega$ is the effective (average) Rabi frequency
of the pulse.}  \label{table1}
\end{table}

The problem becomes more complicated when the physical qubits are not
exactly identical, i.e., when there is inhomogeneity resulting in the
qubits having slightly different resonance frequencies, for instance.
In this case we can switch to frequency-selective addressing, i.e., try
to implement rotations on individual physical qubits by applying
frequency-selective geometric pulse sequences in resonance with each
qubit, either sequentially or concurrently.  However, the pulse
amplitudes (Rabi frequencies) have to be much smaller than the frequency
detuning between the different qubits in this case to avoid off-resonant
excitation effects.  For instance, consider the system
$H[f(t)]=H_0+f(t)H_1$ with
\begin{equation}
  H_0 = -\frac{\hbar}{2}\sum_{n=1}^5 \omega_n \sigma_{z}^{(n)}, \quad
  H_1 =  \frac{\hbar}{2}\sum_{n=1}^5 \sigma_{x}^{(n)},
\end{equation}
where $f(t)=\frac{\gamma_{n}}{2} B_{x}(t)$ and $\sz^{(n)}$ ($\sx^{(n)}$)
denotes a five-fold tensor product, whose $n$th factor is $\sz$ ($\sx$)
and all others are the identity $\ONE$, e.g., $\sz^{(2)}=\ONE \otimes
\sz \otimes \ONE \otimes\ONE \otimes \ONE$.  We can implement a logic
$X$-gate by concurrently applying five Gaussian $\pi$-pulses
\begin{equation}
  f_n(t) = q \sqrt{\pi} e^{-q^2(t-t_n/2)^2} \cos(\omega_n t),
\end{equation}
i.e., choosing $f(t)=\sum_{n=1}^5 f_n(t)$, where $\omega_n$ are the
resonant frequencies $\{6,8,10,12,14\}$ above.  $t_n$ is the length of
the $n$th pulse.  If the pulses have equal lengths and are applied
concurrently then $t_n=t_F$ for all $n$.  We can also apply the pulses
sequentially but this significantly increases the total gate operation
time as $t_F=\sum_n t_n$ in this case, and we will not consider this
case here.

We can quantify the gate fidelity modulo global phases as the overlap of 
the actual gate implemented $U(t_F)$ with the target gate $W$,
\begin{equation}
  \label{eq:fidelity1}
  |\F_W(U(t_F))| = \frac{1}{N} |\Tr[W^\dag U(t_{F})]|,
\end{equation}
where the factor $\frac{1}{N}$, $N=\dim\H$, is a normalization factor
included to ensure that $|\F|$ varies from $0$ to $1$, with unit
fidelity corresponding to a perfect gate.  For the model system above
simulations suggest that we require at least approximately $t_F=440$
time units for Gaussian pulses with $q=0.01$ to achieve $>99.99$\%
fidelity.  

\begin{figure}
\center
\includegraphics[width=0.7\textwidth]{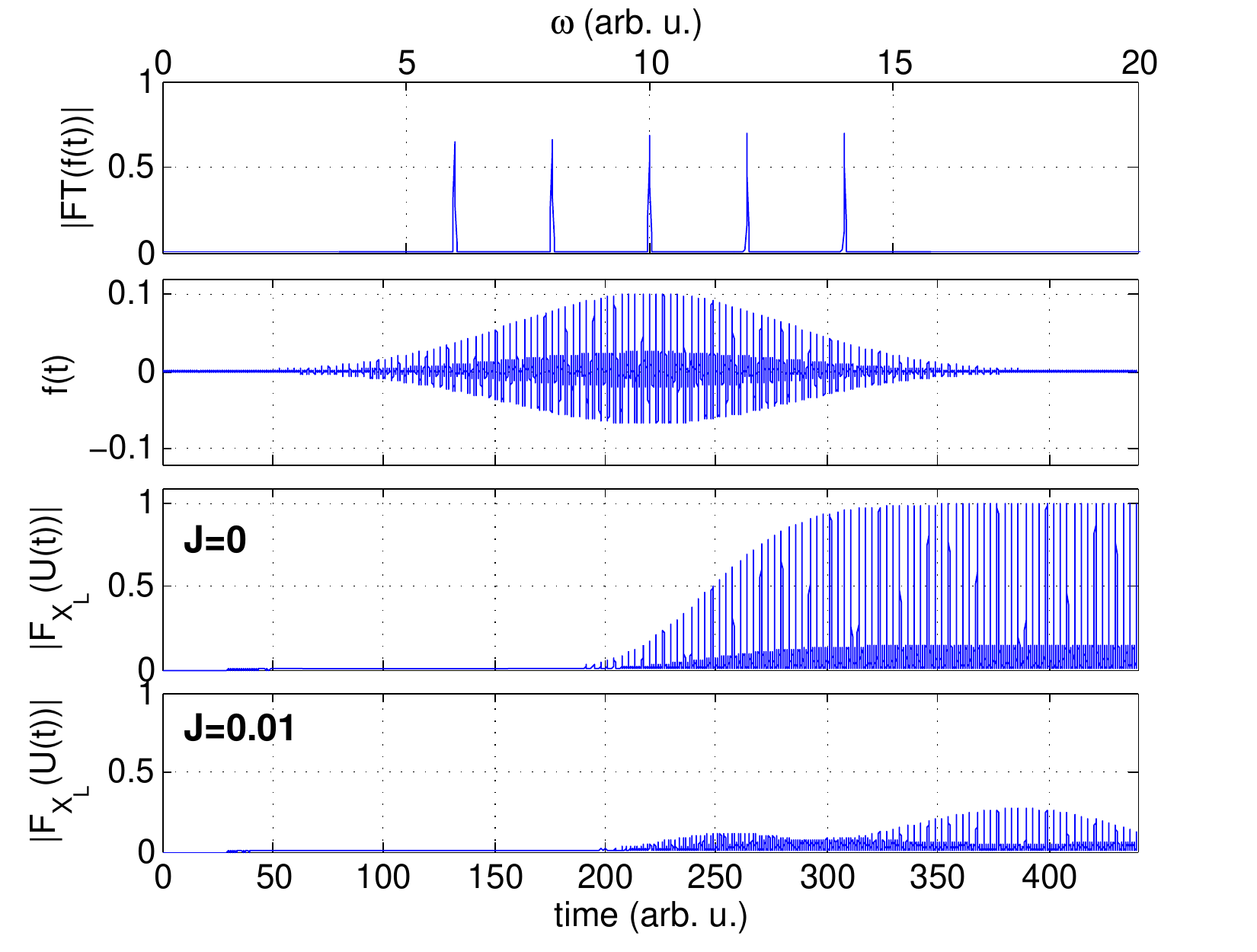}
\caption{A field composed of five concurrent Gaussian $\pi$-pulses with
frequencies $\{6,8,10,12,14\}$ achieves a logic $X$-gate in the absence
of interqubit coupling ($J=0$) but it is ineffective, for $J=0.01$,
achieving a maximum fidelity $|\F|$ of only 27\%.  The high-frequency
oscillations of the fields and fidelity are due to both being computed
and shown in the stationary laboratory frame (no approximations) rather
than in a multiply-rotating frame.}  \label{fig:geom1}
\end{figure}

\begin{figure}
\center
\includegraphics[width=0.7\textwidth]{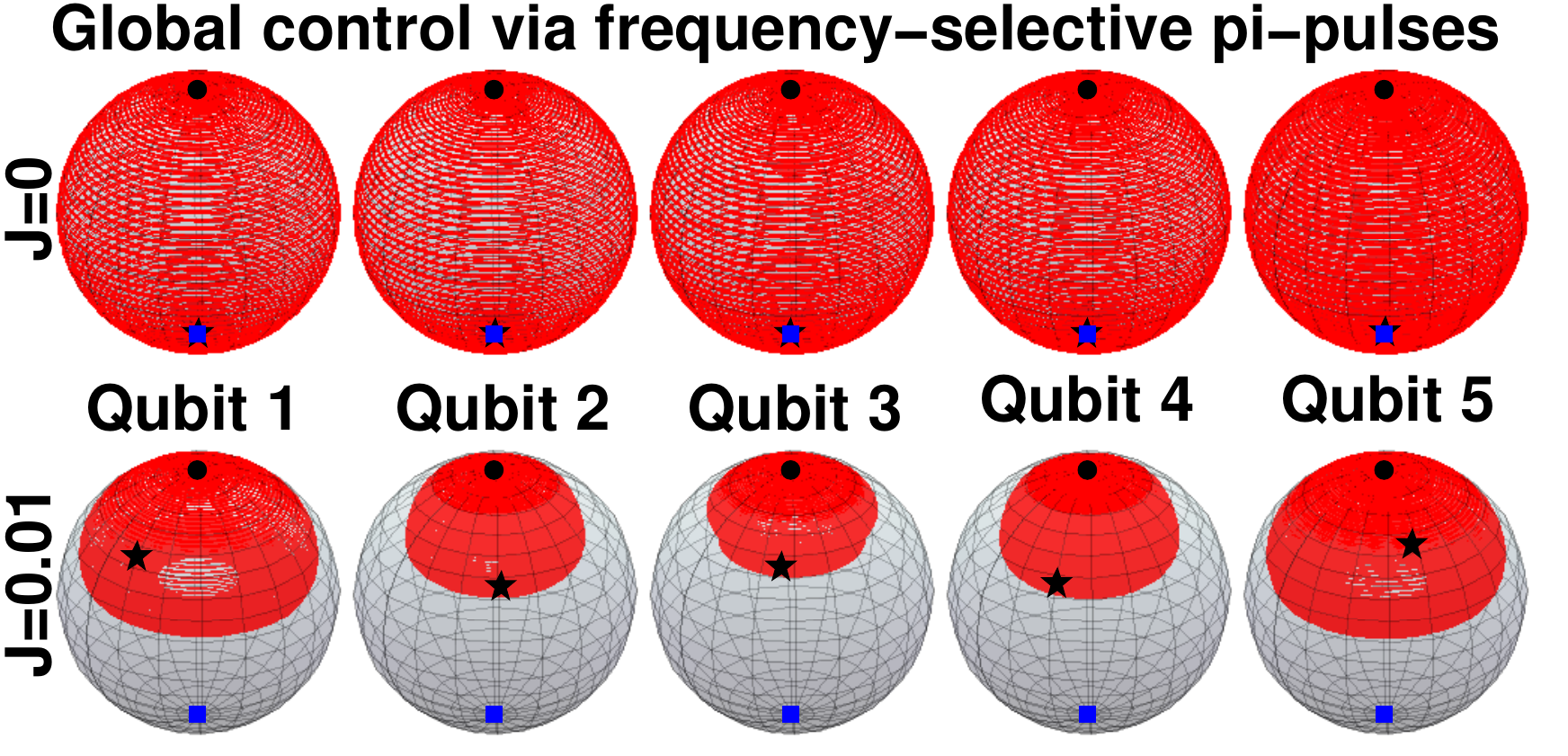}
\caption{Trajectories (red) traced out by the Bloch vectors of the
individual qubits on the Bloch sphere subject to the five concurrent
$\pi$-pulses above for $J=0$ (top) and $J=0.01$ (bottom).  ($\bullet$
initial state, $\bigstar$ final state, $\tiny{\color{blue}\blacksquare}$ target
final state (south pole)} \label{fig:geom2}
\end{figure}

\begin{figure}
\center
\includegraphics[width=0.7\textwidth]{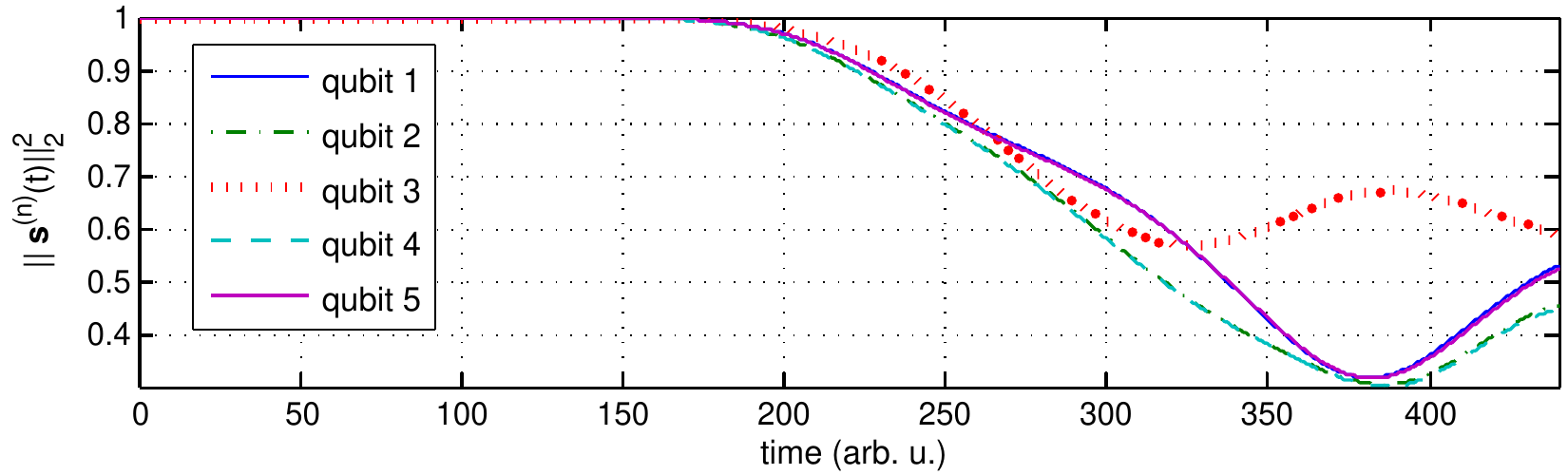}
\caption{Length of the Bloch vectors $\vec{s}^{(n)}(t)$ of the five
physical qubits as a function of time for $J=0.01$.  Its decrease is
evidence of the formation of entanglement between the individual qubits.}
\label{fig:geom3}
\end{figure}

Fig.~\ref{fig:geom1} shows that the frequency-selective pulses perform
as expected if the qubits are non-interacting.  The control performs
poorly, however, if non-zero couplings between adjacent physical qubits
are present, e.g., if we add a simple uniform Ising coupling term
\begin{equation}
  \label{eq:HI}
  H_I = \hbar J \sum_{n=1}^4 \sigma_z^{(n)} \sigma_z^{(n+1)}
\end{equation}
between neighboring qubits, then the fidelity drops from $99.99$\% for
$J=0$ to $99.78$\% for $J=0.0001$, to $86.90$\% for $J=0.001$, to
$26.97$\% for $J=0.01$.  Thus, even if the Ising coupling frequency $J$
is four orders of magnitude smaller than the median transition frequency
$\omega=10$, the simple geometric pulse sequence becomes ineffective.
This can be attributed to the creation of entanglement between the
physical qubits over the duration of the pulse.  The evolution of the
five physical qubits on the Bloch sphere under the five concurrent
$\pi$-pulses in Fig.~\ref{fig:geom2} shows that the presence of the
couplings completely changes the dynamics.  In the absence of couplings
the composite Gaussian pulse achieves the desired simultaneous bit flip
for all the qubits, but in the presence of even moderate Ising couplings
between adjacent qubits the pulse sequence becomes ineffective.  A plot
of the length of the Bloch vectors (Fig.~\ref{fig:geom3}) for $J=0.01$
shows a marked decrease for all qubits, indicating the development of
significant entanglement between the qubits.  Due to the nature of the
couplings (Ising-type) the development of entanglement is not immediate.
If we start in the product state $\ket{00000}$ then the effect of the
couplings is felt only after sufficient coherence has been created by
the pulse.  The product state $\ket{00000}$ was chosen as an initial
state, although it is not a logic state of the stabilizer code, because
it serves well to illustrate the dynamics of the driving system.

\subsection{Local control}

The second control paradigm we consider local control through selective
detuning of individual dots in a quantum register via local voltage
gates, as recently considered, e.g., in~\cite{HILL2005}.  The individual
qubits here are subject to a fixed, global magnetic field $\vec{B}(t)$
as well as local voltage gates that allows us to control the energy
level splitting $\omega_n$ of each of the qubits.  If there is no
coupling between qubits, setting $\vec{B}(t)=B(\cos\omega t,-\sin\omega
t,B_z)$ leads to a Hamiltonian of the form
\begin{equation}
 \label{eq:local} \fl
  H=\frac{\hbar}{2}\sum_{n}\gamma_{n} B[\cos(\omega t)\sigma_{x}^{(n)} 
    -\sin(\omega t)\sigma_{y}^{(n)}]
    +\frac{\hbar}{2}\sum_{n}\omega_{n}(\vec{V}(t))\sigma_{z}^{(n)}
\end{equation}
where $\omega_n(\vec{V})$ indicates that the dependence of the energy
level splitting $\omega_n$ of the $n$th on external control voltages
$\vec{V}$.  It is convenient in this case to transform to a moving frame
$U_0(t)=\exp(it\omega\frac{1}{2}\sum_{n}\sz^{(n)})$ rotating at the
field frequency $\omega$, in which the Hamiltonian takes the form
\begin{equation}
  H = \hbar\Omega\sum_{n}\sigma_{x}^{(n)}
          +\hbar\sum_{n}u_{n}(t)\sigma_{z}^{(n)},
\end{equation}
if we set $\Omega=\frac{1}{2}B\gamma_{0}$ and
$u_{n}(t)=\frac{1}{2}\Delta\omega_{n}(\vec{V}(t))$ with
$\Delta\omega_n(\vec{V}(t))=\omega_n(\vec{V}(t))-\omega$, and for
simplicity assume $\gamma_n=\gamma_0$ for all $n$.  In the following we
do not consider the architecture-specific functional dependence of
resonant frequencies on the gate voltages $\vec{V}$ and cross-talk
issues~\cite{KAND2006,SCHI2007} and take
$\vec{u}(t)=(u_{1}(t),u_{2}(t),\ldots,u_{5}(t))$ to be independent
controls.

\begin{figure}
\center\includegraphics[width=0.9\textwidth]{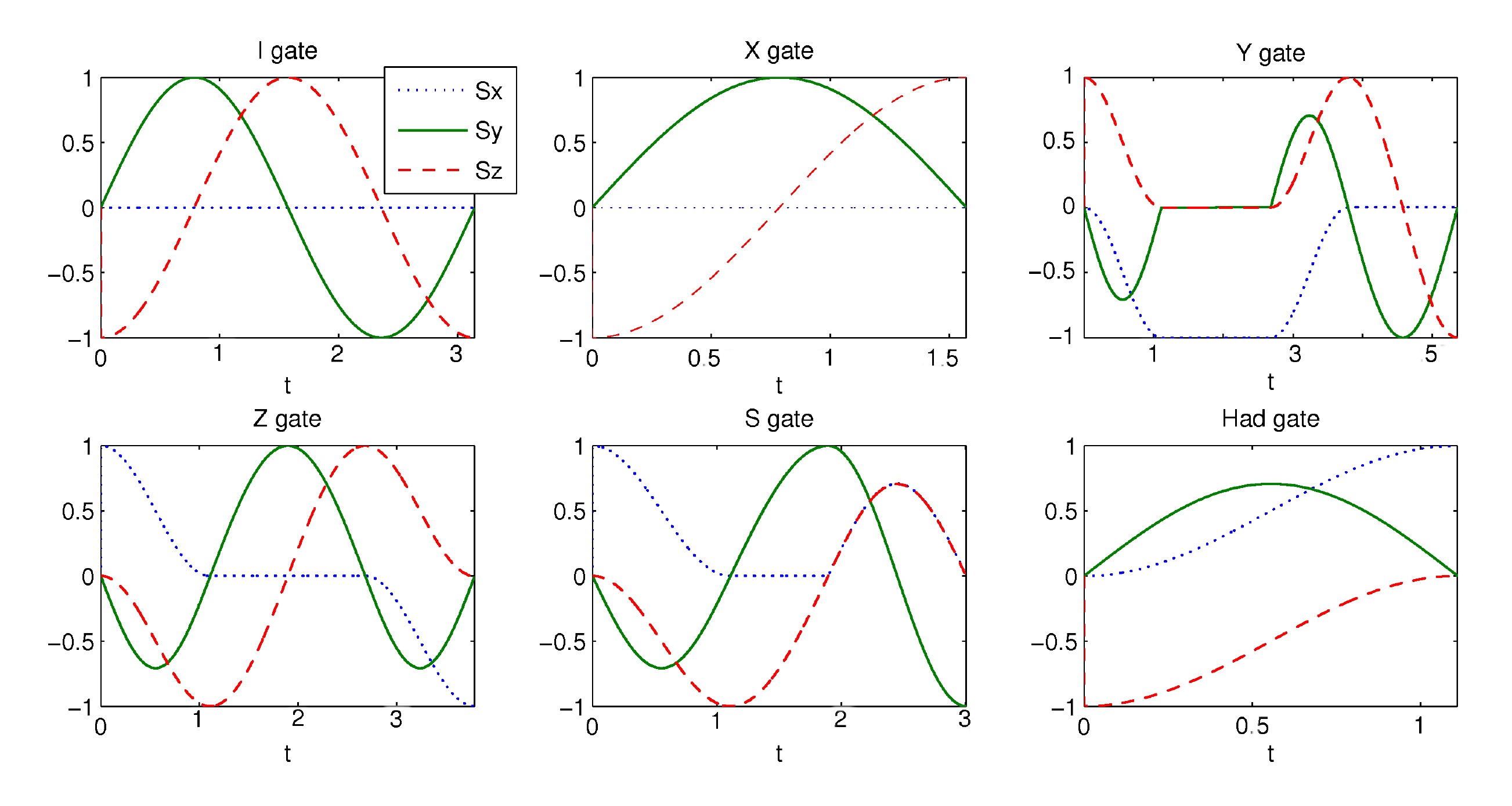}
\caption{Time-evolution of Bloch vector components
 $\vec{s}=(x,y,z)$ for various elementary gates using local
geometric control with time in units of $\frac{\pi}{2}\Omega^{-1}$, 
where $\Omega$ is the Rabi frequency of the fixed global field.}  
\label{fig:geomlocal}
\end{figure}

In the absence of interactions, it is again relatively straightforward
to implement arbitrary rotations on any qubit using a simple geometric
control design.  We can clearly perform simultaneous rotations about the
$\uvec{x}$-axis on all physical qubits simply by choosing the detuning
parameters $u_n(t)=0$.  To rotate qubits about the $\uvec{y}$-axis we can
make use of the following result~\cite{HILL2005}
\begin{equation}
 R_{\uvec{y}}(4\phi)
 =R_{\uvec{x}}(\pi)R_{\uvec{n}}(\pi)R_{\uvec{x}}(\pi)R_{\uvec{n}}(\pi),
\end{equation}
where $R_{\uvec{x}}(\pi)$ is a $\pi$ rotation about the $\uvec{x}$-axis
and $R_{\uvec{n}}(\pi)$ is a $\pi$ rotation about
\begin{equation}
   \uvec{n} = \cos\phi\uvec{x}+\sin\phi\uvec{z},
\end{equation}
where $\cos\phi=\Omega/\Omega_0$ and $\sin\phi=u/\Omega_0$ with
$\Omega_0=\sqrt{\Omega^2+u^2}$ and $u_n(t)=u$.  If we can achieve
detunings of the same magnitude as the fixed coupling, $u=\Omega$, then
$\phi=\frac{\pi}{2}$ and we can therefore implement $\pi$-rotations
about the axis $\uvec{n}=\frac{1}{\sqrt{2}}(1,0,1)^T$, which corresponds
to a Hadamard gate, and rotations about the $z$-axis using
\begin{equation}
  R_{\uvec{z}}(\theta) = \Had R_{\uvec{x}}(\theta) \Had.
\end{equation}
Table~\ref{table2} summarizes the sequences of rotations that implement
the elementary single qubit gates, and Fig.~\ref{fig:geomlocal} shows
the evolution of the components of the Bloch vector for each of gates
assuming piecewise constant controls.  The transversal gates can be
implemented by simply applying the same sequence of rotations on each
qubit in parallel.

An advantage of the local control scheme compared to frequency selective
global control is potentially much faster gate operations because the
gate operation times are determined by the Rabi frequency $\Omega$ of
the global field, which is not limited by the need to avoid off-resonant
excitation, although the need to be able to induce detunings on the
order of $\Omega$ imposes some constraint in practice.  An added bonus
of the shorter gate operations times is that the scheme is far less
sensitive to the presence of small couplings between the qubits.  Adding
a fixed Ising coupling term of the form~(\ref{eq:HI}) does reduce the
gate fidelities, but not nearly as much as was the case for the global
control scheme as the much shorter gate operation times mean that the
qubits have far less time to interact with each other.

A significant disadvantage of this control scheme, however, is that the
implementation times for different gates using the type of simple
geometric pulse sequences described differ significantly, which is
problematic in a scalable architecture where local operations on
different logical units are to be implemented concurrently.  Although
the implementation times in units of the Rabi frequency for different
gates also varied in the global control case, in the former case the
Rabi frequencies of the pulses were variable (within a certain range),
allowing adjustment of the gate implementation times.  This adjustment
is not easily possible in the local control case because the Rabi
frequency $\Omega$ of the global field must remain fixed.

\begin{table}
\begin{center}
\begin{tabular}{|c|c|c|}
\hline 
Gate & Sequence of rotations & Pulse length\\\hline
$X$ & $R_{\uvec{x}}(\pi)$ & 1\\\hline 
$Y$ & $R_{\uvec{x}}(\pi)R_{\uvec{n}}(\pi)R_{\uvec{x}}(\pi)R_{\uvec{n}}(\pi)$
    & $2+\sqrt{2}$\\\hline 
$Z$ & $R_{\uvec{n}}(\pi)R_{\uvec{x}}(\pi)R_{\uvec{n}}(\pi)$ & $1+\sqrt{2}$\\\hline 
$I$ & $R_{\uvec{x}}(2\pi)$ & 2\\\hline 
$S$ & $R_{\uvec{n}}(\pi)R_{\uvec{x}}(\pi/2)R_{\uvec{n}}(\pi)$ 
    & $1/2+\sqrt{2}$\\\hline 
$T$ & $R_{\uvec{n}}(\pi)R_{\uvec{x}}(\pi/4)R_{\uvec{n}}(\pi)$ 
    & $1/4+\sqrt{2}$\\\hline 
$\Had$ & $R_{\uvec{n}}(\pi)$ & $1/\sqrt{2}$\\\hline
\end{tabular}
\end{center}
\caption{Decomposition of elementary gates in terms of rotations about
$\uvec{x}$ and $\uvec{n}=\frac{1}{\sqrt{2}}(1,0,1)^T$ axes and total gate
duration in units of $\frac{\pi}{2}\Omega^{-1}$, where $\Omega$ is the 
effective (average) Rabi frequency of the fixed global field.}  
\label{table2}
\end{table}

In summary, we can in principle implement quantum logic gates on encoded
qubits using simple control pulse sequences based on geometric control
ideas in both the local and global control setting.  However, the simple
schemes have serious shortcomings.  Frequency-selective global control
schemes, for instance, requires comparatively long pulses to minimize
off-resonant excitation.  In the local control setting it is difficult
to implement different gates concurrently on different qubits due to
variable gate operation times.  Both schemes also assume non-interacting
physical qubits, and tend to perform poorly in the presence of fixed
couplings due the creation of entanglement between qubits.  Also, some
essential gates such as the $T$-gate cannot be implemented transversally
at all.  This raises the question if we can overcome such problems using
optimal control, which we will consider in the next sections.

\section{Optimal Control}
\label{sec:OCT}

The problem of finding the control pulses that produce the desired
quantum gates can be treated as an optimization problem.  The procedure
in general involves choosing a measure of the gate fidelity to be
maximized, suitably parameterizing the control fields and finding a
solution to the resulting optimization problem.  If the evolution of the
system is given by a unitary operator $U(t)$ satisfying the Schroedinger
equation
\begin{equation}
 \label{eq:SE}
  \dot{U}(t,t_0) 
  =-\frac{i}{\hbar}\left[H_{0}+\sum_{m=1}^{M}u_{m}(t)H_{m}\right]U(t,t_0),
\end{equation}
where $H_0$ is a drift Hamiltonian and $H_m$, $m=1,\ldots M<\infty$, are
control Hamiltonians, the task is to find a control $\vec{u}(t)$ that
maximizes the gate fidelity
\begin{equation}
  \label{eq:fidelity2}
  \F_W[U(t_F,t_0)] = \frac{2}{N} \Re \Tr[W^\dag U(t_F,t_0)]
\end{equation}
or similar figure of merit for some fixed target time $t_F$
\footnote{The definition of the fidelity~(\ref{eq:fidelity2}) is
stricter than the definition of $|\F_W[U(t_F,t_0)]|$ in
(\ref{eq:fidelity1}), which disregards the global phase of the target
operator.  In general, one does not care about the global phase of the
operator $U$, whence the weaker measure (\ref{eq:fidelity1}) we used
earlier is sufficient.  Although, one can modify the following
derivation to obtain an explicit algorithms to maximize the $|\F_W(U)|$,
the expressions are more complicated due to the absolute value, which is
why we focus on optimizing the stricter fidelity measure $\F_W(U)$,
which takes the global phase into account here.  One could argue that
this is slightly unfair to the optimal control algorithm as we have
actually made the problem harder.  However, since we are working in
$SU(32)$ the only phase factors we are excluding are roots of unity.
Although these can be problematic, they did not seem to pose a problem
for the algorithm in our case, i.e., we had no problem finding solutions
with $W^\dag U$ close to the identity $I$, which is why we did not
consider modifying the algorithm to optimize $|\F_W(U)|$ instead,
although this could be done if necessary.}.

To derive an iterative procedure for finding optimal controls, note that
changing a given control $\vec{u}$ by some amount $\Delta\vec{u}$
changes the corresponding propagator (see \ref{appendix:A})
\begin{eqnarray}
  \Delta U_{\mathbf{u}}(t,t_0) 
  &\equiv&  U_{\vec{u+\Delta u}}(t,t_0)-U_{\vec{u}}(t,t_0)\nonumber\\
  &=& -\frac{i}{\hbar}\int_{t_0}^{t}\sum_{m}\Delta u_{m}(\tau)
      U_{\vec{u}}(t,\tau)H_m U_{\vec{u}+\vec{\Delta u}}(\tau,t_0)\, d\tau.
  \label{eq:DeltaU}
\end{eqnarray}
The corresponding change in the fidelity is 
\begin{eqnarray}
 \Delta \F(t_F) 
 &=& \F_{\mathbf{u+\Delta u}}(t_F)-\F_{\mathbf{u}}(t_F) \nonumber\\
 &=& \frac{2}{N}\Re\Tr[W^{\dag}(U_{\vec{u+\Delta u}}(t_F,t_0)
     -U_{\vec{u}}(t_F,t_0))] \\
 &=& \frac{2}{N\hbar}\sum_{m}\int_{t_{0}}^{t_F}\!\!\!\! \Delta u_{m}(\tau)\Im
     \Tr[W^{\dag}U_{\vec{u}}(t_F,\tau)H_{m}U_{\vec{u+\Delta u}}(\tau,t_0)]\, d\tau.
 \nonumber
\end{eqnarray}
This shows that setting
\begin{equation}
 \label{eq:deltau}
 \Delta u_{m}(t) = \epsilon_m(t)\Im\Tr[W^{\dag}U_{\vec{u}}(t_{F},t)H_m
 U_{\vec{u+\Delta u}}(t,t_0)]\label{eq:update1}
\end{equation}
with $\epsilon_{m}(t)>0$ for all $t$ will increase the fidelity at the
target time $t_F$.  Thus, a basic algorithm for maximizing $\F$ is to
start with some initial guess $\vec{u}_{0}(t)$ and solve the Schroedinger
equation iteratively while updating the control according to the rule
\begin{equation}
  \vec{u}^{(n+1)} =\vec{u}^{(n)} + \Delta\vec{u}^{(n)},
\end{equation}
with $\Delta\vec{u}^{(n)}$ as in~(\ref{eq:update1}).  Note that this is
an implicit update rule as the RHS of (\ref{eq:deltau}) depends on
$\Delta\vec{u}(t)$, and the search space is infinite dimensional,
consisting of functions on defined on an interval of the real line.

To derive useful practical algorithms the controls $u_{m}(t)$ need to be
discretized, and a simplified explicit update rule is desirable.  The
simplest and most common approach to discretization is to approximate
the continuous fields by piecewise constant functions, i.e., we divide
the total time $t_{F}$ into $K$ steps, usually, though not necessarily,
of equal duration $\Delta t=t_F/K$, and take the control amplitudes to
be constant during each interval $I_k=[t_{k-1},t_k)$.  We then have
\begin{equation}
  \F_W[U(t_F)] = \frac{2}{N} \Re \Tr\left[ 
   W^\dag U_{\vec{u}}^{(K)} U_{\vec{u}}^{(K-1)} \cdots 
   U_{\vec{u}}^{(2)} U_{\vec{u}}^{(1)} \right],
\end{equation}
where the step propagator for the $k$th step is given by
\begin{equation}
  \label{eq:Uk}
  U_{\vec{u}}^{(k)} = \exp\left[-\frac{i}{\hbar}\Delta t
        \left(H_{0}+\sum_{m=1}^{M} u_{mk} H_{m} \right)\right]
\end{equation}
and $u_{mk}$ is the amplitude of the $m$th control field during the
$k$th step.  Furthermore, if we change the control field by $\Delta
u_m(t)=\Delta u_{mk}$ for $t\in I_k$ then the change in the fidelity
at the target time $t_F$ is (see \ref{appendix:B})
\begin{equation}
 \label{eq:DeltaF}
 \fl \Delta\F(t_F) =
   \frac{2}{N\hbar}\sum_{m} \Delta u_{mk} 
      \Im\Tr \left[ W^\dag 
      U_{\vec{u}}^{(K)}\cdots U_{\vec{u}}^{(k+1)} 
      \Delta U_m^{(k)} 
      U_{\vec{u}+\Delta\vec{u}}^{(k-1)} \ldots 
      U_{\vec{u}+\Delta\vec{u}}^{(1)}\right]
\end{equation}
with $U_{\vec{u}}^{(k)}$ and $U_{\vec{u}+\Delta\vec{u}}^{(k)}$ as 
defined in (\ref{eq:Uk}) and 
\begin{equation}
 \label{eq:DeltaUmk}
 \Delta U_m^{(k)} = \int_{t_{k-1}}^{t_{k}}\!\!\! 
        U_{\vec{u}}(t_k,\tau)H_{m}U_{\vec{u+\Delta u}}(\tau,t_{k-1})]\, d\tau.
\end{equation}
This shows that the fidelity will increase at time $t_F$ if we change
the field amplitude $u_{mk}$ in the time interval $I_k$ by
\begin{equation*} 
   \Delta u_{mk} = \eps_{mk} \Im \Tr \left[
       W^\dag U_{\vec{u}}^{(K)} 
      \cdots U_{\vec{u}}^{(k+1)} 
      \Delta U^{(k)}(t) U_{\vec{u}+\Delta\vec{u}}^{(k-1)} \ldots 
      U_{\vec{u}+\Delta\vec{u}}^{(1)}\right]
\end{equation*}
for $\eps_{mk}>0$.  This is still an implicit update rule as $\Delta
U^{(k)}$ on the RHS depends $\Delta u_{mk}$.  However, when $\Delta
t=t_k-t_{k-1}$ and $\Delta u_{mk}$ are sufficiently small then we can
approximate $\Delta U_m^{(k)}$, e.g., setting $\Delta U_m^{(k)}\approx
U_{\vec{u}}^{(k)} H_m$, which yields the familiar explicit update rule
\begin{equation} 
\label{eq:deltau_k}
  \Delta u_{mk}=\eps_{mk} \Im\Tr \left[ W^\dag U_{\vec{u}}^{(K)} 
      \cdots U_{\vec{u}}^{(k)} H_m U_{\vec{u}+\Delta\vec{u}}^{(k-1)} 
      \ldots U_{\vec{u}+\Delta\vec{u}}^{(1)}\right].
\end{equation}
In practice we can now solve the optimization problem by starting with
an initial guess for the controls $\vec{u}_m^{(0)}$, calculating
$U^{(0)}(t_F)$ and then integrating alternatingly backwards and forwards
while updating the controls in each time step according to the explicit
update rule (\ref{eq:deltau_k}) until no further improvement in the
fidelity is possible.  Ignoring errors to due numerics (e.g. limited
precision floating point arithmetic, etc), if the algorithm is
implemented strictly as presented and the $\eps_{mk}$ are chosen in each
time step and iteration such that the value of the fidelity at the final
time does not decrease, then convergence is guaranteed as the fidelity
is nondecreasing and bounded above.  However, we cannot guarantee that
the fidelity will converge to the global maximum, especially when there
are constraints on the field amplitudes or time resolution $\Delta t$ of
the fields.  The algorithm will stop when it is no longer possible to
change $u_{mk}$ for any $m$ or $k$ such as to increase the final
fidelity.

The update rule~(\ref{eq:deltau_k}) is similar to the update rule of the
familiar GRAPE algorithm~\cite{KHAN2005}, but in the GRAPE algorithm the
update is global, i.e., the field amplitudes $u_{mk}$ for all times
$t_k$ are updated concurrently in each iteration step based on the
propagators $U^{(k)}_{\vec{u}}$ using the fields from the previous
iteration step, which is equivalent to setting
\begin{equation} 
\label{eq:update_grape}
  \Delta u_{mk}=\eps_{m} \Im\Tr \left[ W^\dag U_{\vec{u}}^{(K)} 
      \cdots U_{\vec{u}}^{(k)} H_m U_{\vec{u}}^{(k-1)} 
      \ldots U_{\vec{u}}^{(1)}\right]
\end{equation}
for all $m$ and $k$.  This global update has certain advantages such as
easy parallelizability, but it requires the field changes in each step
to be small to maintain the validity of the approximations involved and
ensure an overall increase in the fidelity.  The parameter $\eps_m$ is
critical in this regard and must be chosen carefully to ensure monotonic
convergence.  One way this can be achieved is by choosing $\eps_m$ to be
very small (steepest decent) but this comes at the expense of very slow
convergence.  In practice one would therefore usually employ step-size
control in the form of linesearch, quasi-Newton methods or conjugate
gradients to accelerate convergence.

Despite the similarities in the update rules, the sequential local
update method is quite different.  With sequential local update the
fidelity increases in every time step, not just for every iteration, and
we can make much larger changes to the field(s) in each step, with the
magnitude of the allowed changes limited mainly by explicit constraints
on the field amplitudes and the approximations involved in transforming
the implicit update rule to an explicit one, which could be avoided or
improved, and discretization errors.  The sequential local update rule
also allows us to easily enforce certain local constraints.  For
example, a constraint on the field amplitudes of the form $|u_m(t)|<C$
can be trivially incorporated as we simply have to choose $\eps_{mk}$ in
each time step $k$ so that $|u_{mk}+\Delta u_{mk}|\le C$.  Assuming the
initial controls is choosen such that the constraints are satisfied,
$|u_{mk}|\le C$, this is always possible, if necessary by setting
$\eps_{mn}=0$ for this time step.  In this case there will, of course, 
be no increase of the final fidelity during this time step but we will
generally be able to continue to increase the fidelity in the next time
steps, although such an amplitude constraint may eventually prevent us
from increasing the fidelity further for all time steps, in which case
we could be left with a solution for which final fidelity is less than
its optimum value.  Of course, amplitude constraints can also be
incorporated in global update schemes, albeit not quite as trivially.
Another perhaps more interesting feature of the sequential update 
method is that it enables us use variable time steps to increase the
fidelity further should a given time resolution of the fields prove
insufficient to achieve high-enough fidelities.  

The sequential local update method is similar to a class of methods
often referred to as the Krotov method, based on work by
Krotov~\cite{KRO1983,KRO1999}, adapted to quantum systems by Tannor et
al~\cite{TAN1992}, although our formulation is for unitary operators
rather than quantum states and there is no penalty on field energy.  

\section{Optimal Control Implementation of Encoded Logic Gates}
\label{sec:applic}

We now apply optimal control algorithm described in the previous section
to the problem of implementing fault-tolerant logic gates for the
five-qubit stabilizer code.  In particular, we wish to implement a full
set of elementary logic gates in a fixed amount of time for systems
subject to both inhomogeneity and interactions between adjacent qubits.
For the global control system we choose the total Hamiltonian
\begin{equation} \fl
  \label{eq:sys1}
  H^{(\rm g)}[\vec{u}(t)] = 
      -\frac{1}{2}\sum_{n=1}^5 \omega_n \sigma_{z}^{(n)} 
       +J \sum_{n=1}^4 \sigma_z^{(n)} \sigma_z^{(n+1)}
       +u_1(t) \sum_{n=1}^5 \sx^{(n)}
       +u_2(t) \sum_{n=1}^5 \sy^{(n)} \quad
\end{equation}
with $\omega_n\in\{6,8,10,12,14\}$, and the aim is to optimize the two
fields $u_1(t)$ and $u_2(t)$, which roughly correspond to the $x$ and
$y$ components of an external electromagnetic field, to implement the
desired encoded logic gates.  For the local control case we choose the
Hamiltonian
\begin{equation}
  \label{eq:sys2}
  H^{(\rm l)}[\vec{u}(t)] = \Omega\sum_{n}\sx^{(n)}
          + J\sum_{n=1}^4 \sigma_z^{(n)} \sigma_z^{(n+1)}
          + \sum_{n}u_{n}(t)\sz^{(n)}
\end{equation}
where $u_n(t)$ for $n=1,2,3,4,5$ are controls that induces local
detunings and $\Omega=10$ is the Rabi frequency of the fixed external
coupling field.  Here we have chosen units such that $\hbar=1$.

In order to find the optimal solutions using the algorithm outlined
above we must choose the number of time-steps $K$ and the total time
$t_{F}$ and a suitable initial control $\vec{u}^{(0)}$.  To ensure that
the gates are fast and do not require controls with extremely high time
resolution, we aspire to find solutions for $t_F$ and $K$ small.
Unfortunately, there are no simple rules for choosing $t_F$ and $K$.
With some experimentation we were able to find controls that achieved
gate fidelities of approximately $99.99$~\% for the global control
example with $J=1$ for $t_F=125$ with $K=1250$ time steps.  For the
local control example $t_F=30$ and $K=300$ proved sufficient.  The gate
operation times achieved using optimal controls in the global control
setting, although much longer than for the local controls, are still
favorable compared to frequency-selective geometric control pulses, for
which even the $X$-gate required approximately $500$ time units for 
$J=0$ and the method failed for $J$ greater than $10^{-4}$.  Although
the local gates for $\Omega=10$ are much slower than what is
theoretically possible using local geometric control in the absence of
coupling, the optimal controls have the distinct advantage that they
enable us to implement all gates in a fixed amount of time and deal with
non-zero $J$-couplings.  Also, the implementation of the $T$-gate, which
can not be implemented transversally, presents no greater challenge for 
optimal control than any of the other gates.  In fact, the simulations
suggest that the most difficult gate to implement in the global control
setting is the $Y$-gate.  In the local control setting there appeared to
be little difference in the difficulty of implementing different gates.

\begin{figure}
\center
\includegraphics[width=0.7\textwidth]{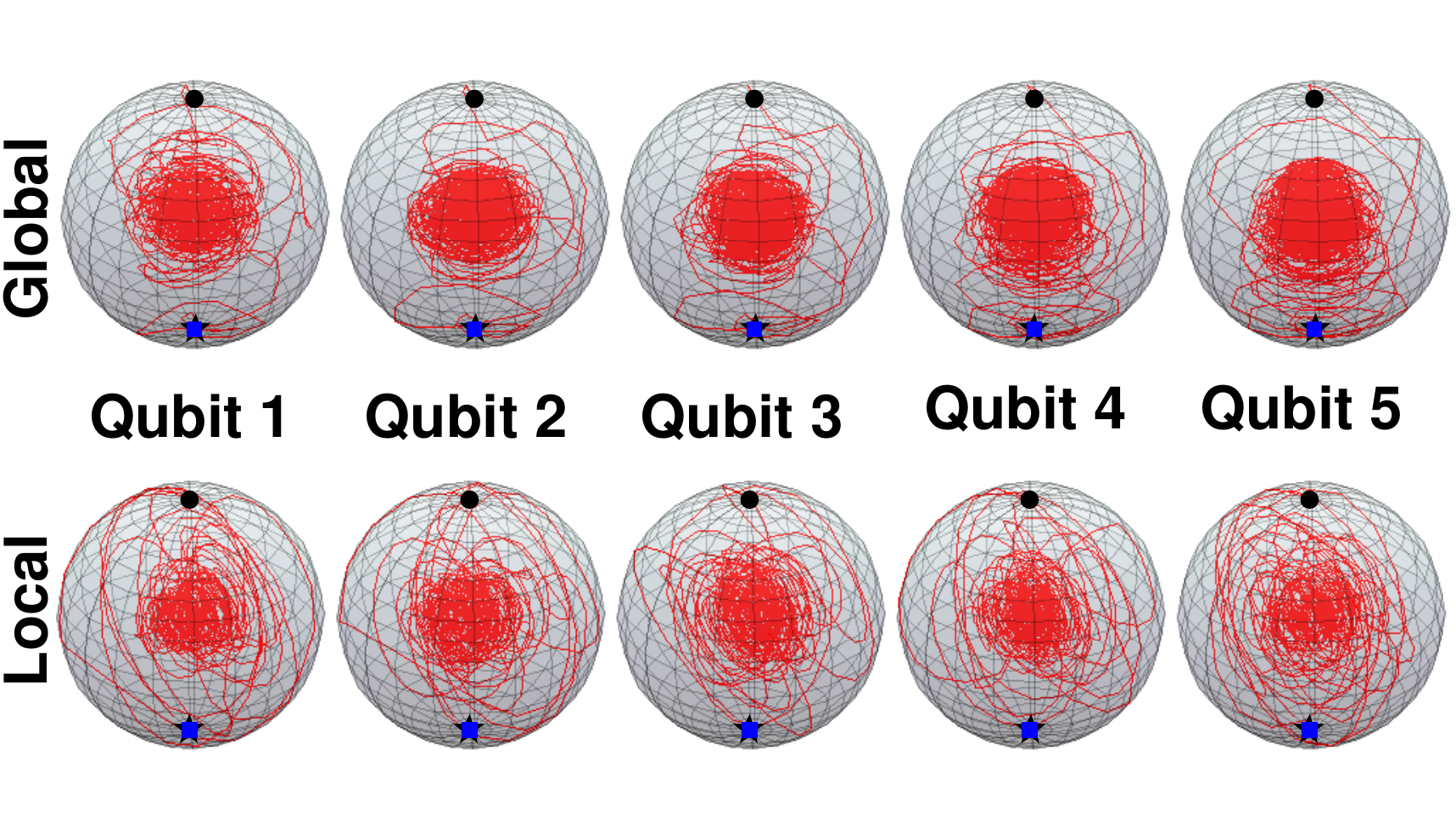}
\caption{Evolution of physical qubits initialized in the state 
$\ket{00000}$ subject to the optimal controls required to implement
the $X$-gate.}
\label{fig:OptX}
\end{figure}

For comparison with the geometric control results, Fig.~\ref{fig:OptX}
shows the trajectories of the physical qubits subject to the optimal
controls for the logic $X$-gate.  Again, the qubits were assumed to be
initialized in the product state $\ket{00000}$.  This is not a logic
state but allows us to easily check visually that the gate effects the
desired simultaneous bit flip on all physical qubits even in the
presence of significant inter-qubit couplings ($J=1$).  The single-qubit
trajectories start out at the surface (north pole) of the Bloch sphere,
corresponding to the product state $\ket{00000}$.  They descend inside
the Bloch ball as the qubits then become entangled, and the individual
qubits follow different paths, but at the final time the trajectories of
all physical qubits converge to the south pole of the Bloch sphere, as
desired.

\begin{figure}
\center\includegraphics[width=\textwidth]{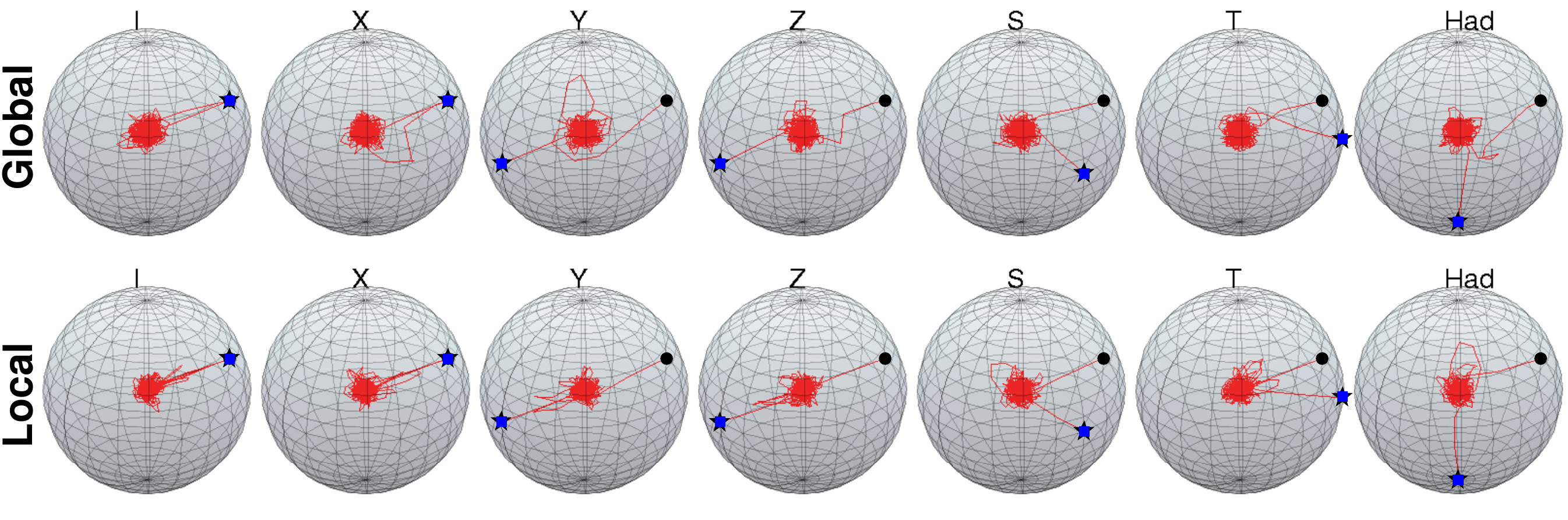}
\caption{Controlled evolution of the initial state
$\ket{\Psi(0)}=(\ket{0_L} +\ket{1_L} )/\sqrt{2}$ projected onto the 2D
subspace spanned by $\ket{0_L} $ and $\ket{1_L} $ for different logic
gates for optimal control examples with global and local control.}  
\label{fig:trajctL2}
\end{figure}

To get an idea of how the optimal controls perform for the other gates
beyond the gate fidelity, Fig.~\ref{fig:trajctL2} shows the projection
of the controlled evolution onto the two-dimensional subspace of the
Hilbert space spanned by the logic states $\ket{0_L} $ and $\ket{1_L} $
for the initial state
$\ket{\Psi(0)}=\frac{1}{\sqrt{2}}(\ket{0_L} +\ket{1_L} )$.  The Bloch
vector was defined to be
\begin{eqnarray}
 s_x(t)&=& 2\Re\left[\ip{0_L}{\Psi(t)}\ip{\Psi(t)}{1_L} \right]\\
 s_y(t)&=&-2\Im\left[\ip{0_L}{\Psi(t)}\ip{\Psi(t)}{1_L} \right]\\
 s_z(t)&=& |\ip{1_L}{\Psi(t)}|^2-|\ip{0_L}{\Psi(t)}|^2,
\end{eqnarray}
where $\ket{\Psi(t)}=U(t)\ket{\Psi(0)}$ and $U(t)$ is the full $32\times
32$ unitary operator describing the evolution of the system under the
given optimal control.  The plots show that the evolution of the system
is very complicated.  Although the initial state is in the 2D subspace
spanned by the logic states $\ket{0_L} $ and $\ket{1_L} $, the fact that
the trajectories (red) are concentrated in the interior of the Bloch
ball indicates that the system spends most of the time outside the
two-dimensional logic subspace, but returns (mostly) to this subspace
near the final time.  Although the gate fidelities as defined in
(\ref{eq:fidelity2}) are about $99.99$~\%, there are small but noticeable
differences between the actual final states ($\bigstar$) and the final
states (blue squares) for an ideal gate.  This is due to the fact that
the normalized gate fidelity for high-dimensional gates is not a very
strict measure as
\begin{eqnarray}
   \norm{W-U(t_F)}_2 &= \sqrt{\Tr[(W-U)^\dag (W-U)]} \\
                     &=\sqrt{2N[1-\F_W(U)]},\nonumber
\end{eqnarray}
with $\F_W(U)$ as defined in Eq.~(\ref{eq:fidelity2}), i.e., the
distance between the operators with respect to the Hilbert-Schmidt norm
is significantly larger than the gate error defined in terms of the
fidelity $1-\F_W[U(t_F)]$.  For $N=32$ a gate fidelity of $>99.99$~\%
allows a Hilbert-Schmidt distance between the operators up to $0.08$.
The distance between the operators with respect to the operator norm
$\norm{X}$ (given by the largest singular value of $X$) is usually
smaller than the Hilbert Schmidt distance but still significantly larger
than the fidelity error.  A summary of different gate error measures for
all gates is given in Table~\ref{table:errors}.

\begin{figure}
\includegraphics[width=\textwidth]{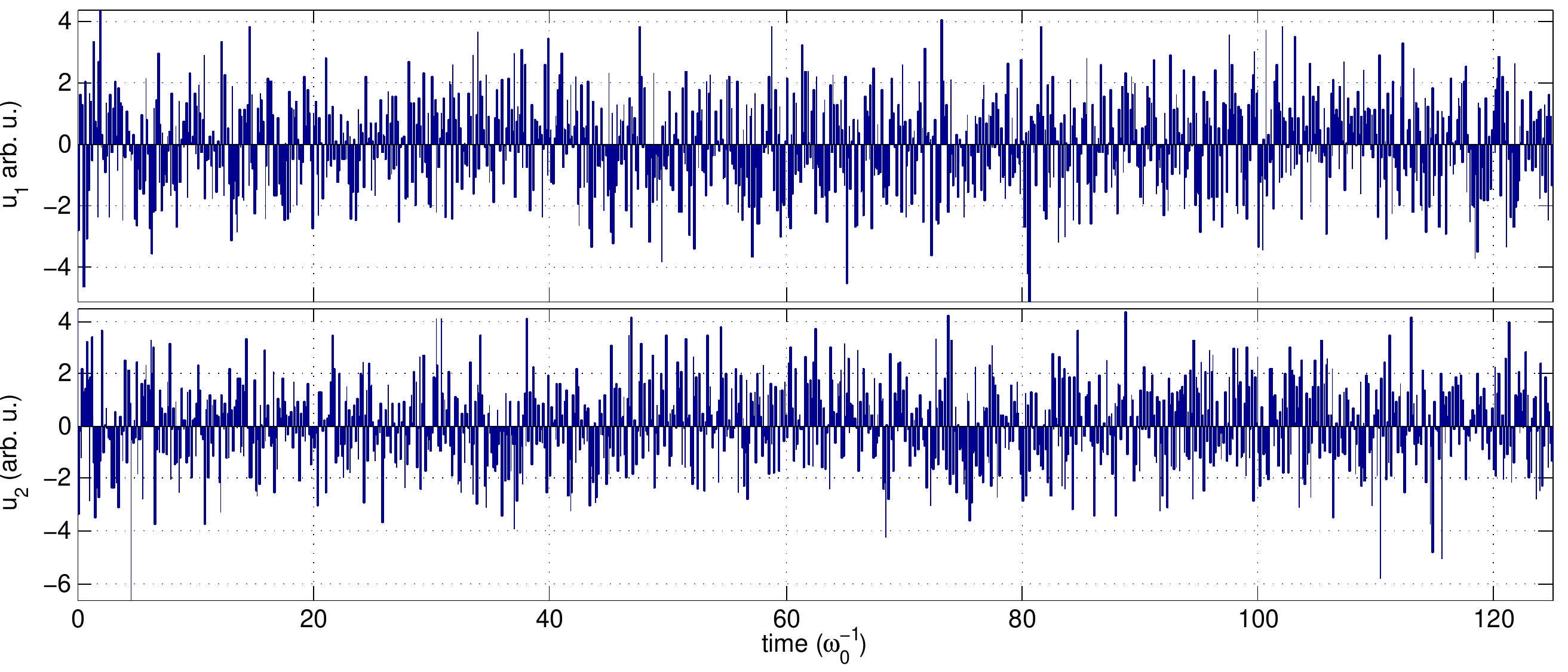} 
\caption{Example of optimal control solution to implement the 
non-transversal $T$-gate for the five qubit stabilizer code using two
independent global fields for system~(\ref{eq:sys1}).}
\label{fig:fields:global}
\end{figure}

\begin{figure}
\rotatebox{90}{\includegraphics[height=\textwidth]{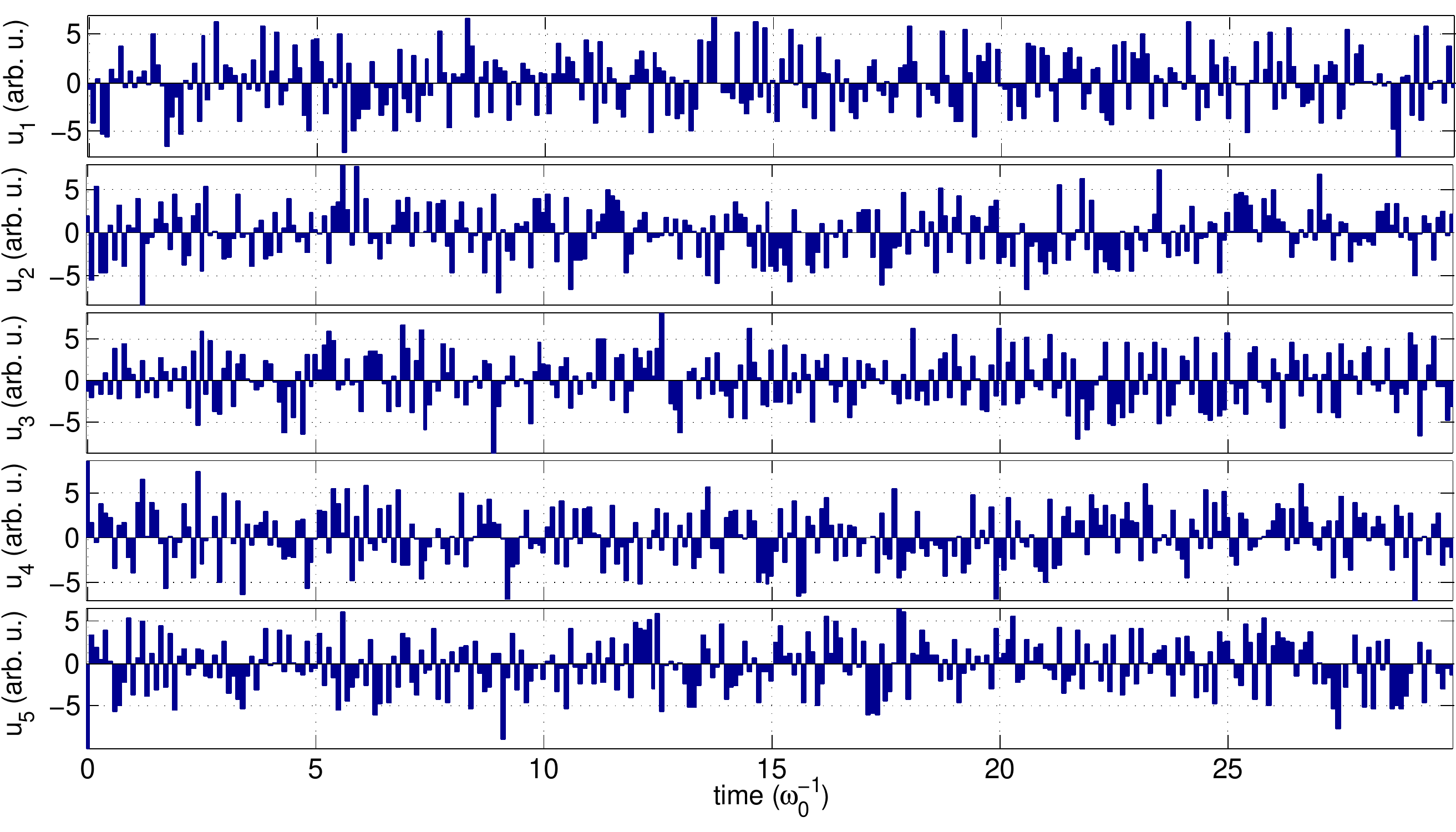}}
\caption{Example of optimal control solution to implement the 
non-transversal $T$-gate for the five qubit stabilizer code using five
independent local detuning fields for system~(\ref{eq:sys2}).}
\label{fig:fields:local}
\end{figure}

Examples of the actual control fields obtained in both the local and
global control setting are shown in Figures \ref{fig:fields:global} and
\ref{fig:fields:local} for the $T$-gate.  The control fields appear
complicated, especially in the global control case.  The local control
fields appear to be better behaved but this is is in part due to the
scaling of the time axis --- the global control pulses are much longer
and thus appear much more compressed when scaled to fit the width of the
text.  Nonetheless the controls are complicated and have a relatively
high bandwidth.  To some extent this is unsurprising considering the
complexity of the system, control goals, and constraints imposed on the
controls in terms of pulse lengths and time resolutions.  However, there
are many possible controls that result in the same gate, corresponding
to alternative trajectories in $SU(N)$ connecting the identity and the
target gate.  A limited number of numerical experiments performed with
different choices for the initial fields suggest that these lead to
different solutions for the controls, but generally the solutions appear
to have similar complexities.  However, it may be possible to exploit
the non-uniqueness of the solutions to find better solutions, either by
changing the parameterization of the fields or imposing carefully chosen
additional penalty terms, which we will consider in future work.

\begin{table}
\center
\begin{tabular}{|l||l|c|c|c|c|c|c|c|} \hline
       &           & I   & X   & Y   & Z   & S   & T   & Had\\\hline\hline
global 
&$\F_W(U)$     &0.9999&0.9999&0.9996&0.9999&0.9999&0.9999&0.9999\\ 
&$\norm{W-U}$  &0.0267&0.0268&0.0573&0.0259&0.0255&0.0261&0.0266\\
&$\norm{W-U}_2$&0.0800&0.0800&0.1644&0.0800&0.0800&0.0800&0.0800\\
&$\norm{W-U}_{\max}$&0.0061&0.0069&0.0163&0.0074&0.0064&0.0071&0.0065\\\hline
local 
&$\F_W(U)$     &0.9999&0.9999&0.9999&0.9999&0.9999&0.9999&0.9999\\
&$\norm{W-U}$  &0.0268&0.0266&0.0266&0.0253&0.0252&0.0250&0.0263\\
&$\norm{W-U}_2$&0.0799&0.0798&0.0799&0.0798&0.0798&0.0798&0.0799\\
&$\norm{W-U}_{\max}$&0.0056&0.0069&0.0069&0.0058&0.0069&0.0068&0.0066\\\hline
\end{tabular}
\caption{Different performance measures for the gates implemented using
optimal control: Fidelity, operator distance $\norm{W-U}$ (largest 
singular value of $W-U$), Hilbert-Schmidt distance 
$\norm{W-U}_2=\sqrt{\Tr[(W-U)^\dag (W-U)]}$, 
and maximum distance between matrix elements $\norm{W-U}_{\max} = 
\max_{m,n}|W_{mn}-U_{mn}|$, where $U=U(t_F)$ and $W$ is the target 
gate.}
\label{table:errors}
\end{table}

\begin{figure}
\center\includegraphics[width=0.49\textwidth,height=1.4in]{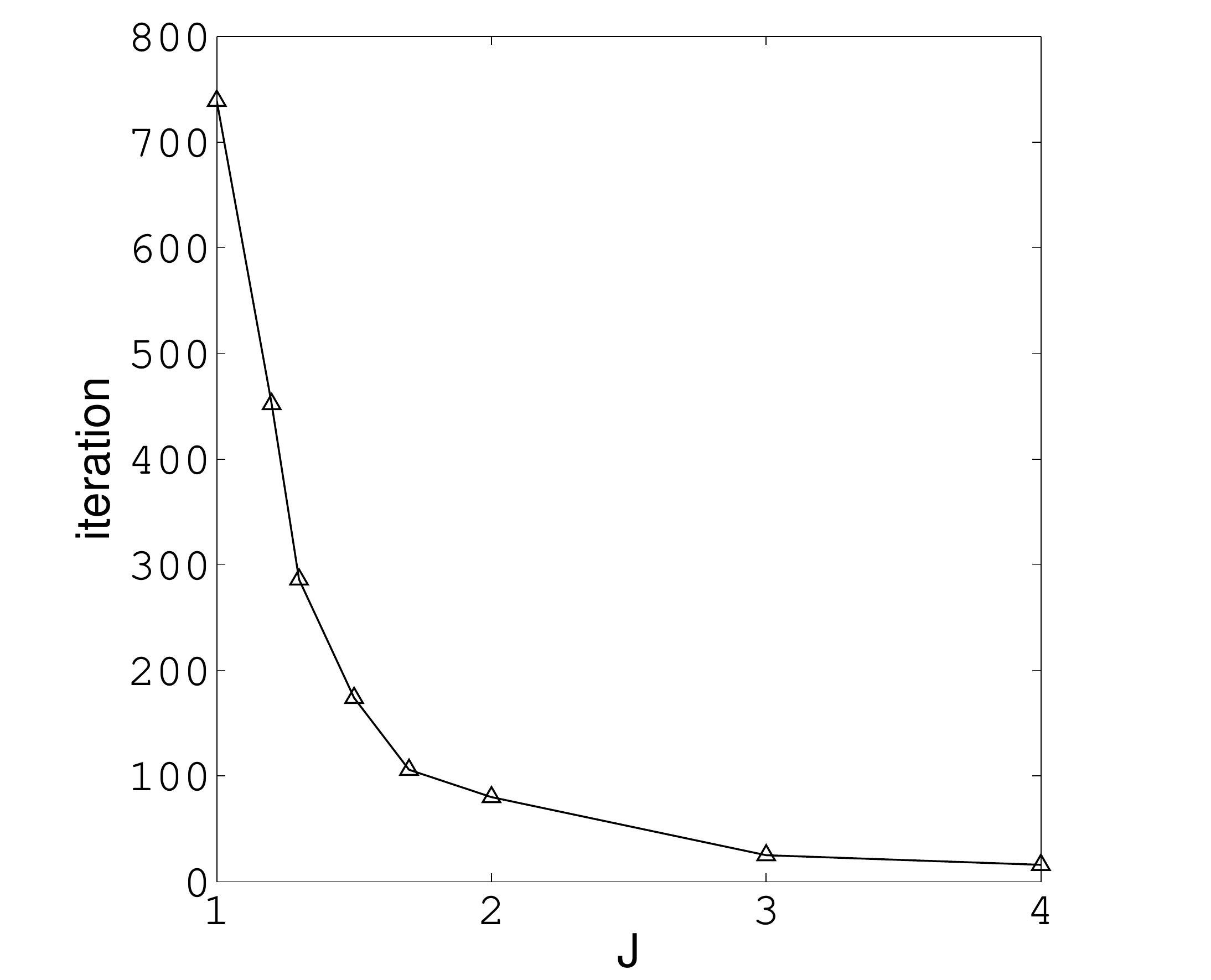} 
\includegraphics[width=0.49\textwidth,height=1.4in]{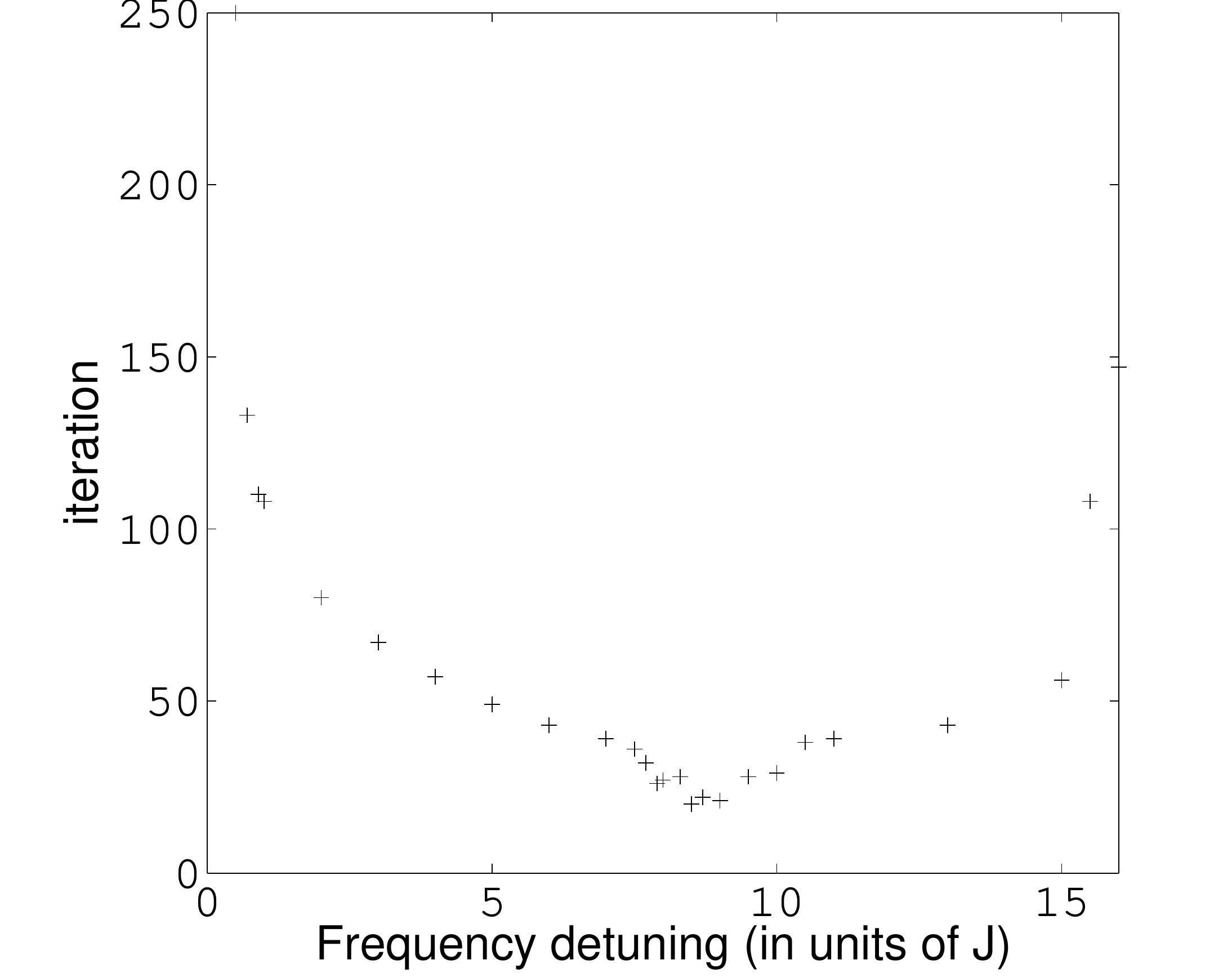}
\caption{Number of iterations required to find solutions with
$\F_W(U)\ge 0.9999$ for (a) $t_{F}=10$, $M=80$ and $\Delta\omega=2$;
(b) $t_{F}=10$, $M=80$ and $J=2$.} 
\label{fig:iterations}
\end{figure}

The presence of fixed couplings between the qubits, which was highly
detrimental for the geometric control schemes, does not appear to be a
problem in the optimal control case, and indeed some of our simulations
suggested that larger $J$-couplings may in fact make it easier to find
optimal control solutions.  This apparently counter-intuitive behavior
prompted us to consider the effect of varying the interqubit coupling
strength $J$ and the detunings $\Delta\omega$.  We can get an estimate
of the difficulty of finding solutions by setting $t_F$ and $K$ to
constant values and comparing the number of iterations required to
archive $99.99$\% fidelity for various values of $J$ and $\Delta\omega$.
A small number of iterations generally implies that it is easy to find a
solution and that solutions may exist for significantly shorter times
$t_F$.  To reduce the computational overhead, we considered the same
model systems but with only three instead of five qubits, and taking as
target gates the standard gates for the three-qubit bit-flip code with
code words $\ket{0_L} =\ket{000}$ and $\ket{1_L}=\ket{111}$.  For this
system the run time of the optimization routine reduces from several
hours for the five qubit code to a few minutes on a single 3GHz CPU.  It
proved easy to find solutions for the system subject to global control
with $J=1$ for $t_F=10$ and $K=80$, so we selected these values as
defaults.  Fig.~\ref{fig:iterations}(a) shows the number of iterations
for different values of $J$.  Although the number of iterations required
for convergence maybe an oversimplified measure of the complexity, the
data indeed suggests that increasing $J$ makes it easier for the
algorithm to find solutions, suggesting that optimal control allows us
to exploit couplings that are usually considered undesirable to speed up
gate operations, especially in the global control setting.  The effect
of changing $\Delta\omega$ is more subtle.  If $\Delta\omega$ is very
small then the selectivity of the pulses is very weak and the algorithms
struggles to find solutions unless $t_F$ is increased.  If
$\Delta\omega$ is very large then it is also hard to find the solutions,
partly because the system can not use off-resonant excitations as much,
but more likely because the bandwidth of the fields available may not be
sufficient to cover the required frequency range.

\section{Conclusion}

We have considered the problem of implementing quantum logic gates
acting on encoded qubits, As a particular example, the five-qubit stabilizer
code was considered.  Despite the complicated encoding, most elementary
logic gates can in principle be implemented fault-tolerantly by applying
the same local gates to all physical qubits.  Thus, it is usually argued
that the implementation of logic gates on encoded qubits can be reduced
to the implementation of single and two-qubit gates on physical qubits,
and that the latter can in principle be implemented using simple pulse
sequences based geometric control ideas.  However, there are several
problems with this.  Not all gates can be implemented transversally.
For the five-qubit code, for instance, it is not possible to implement a
$T$-gate this way, which is an essential gate for universal quantum
computation.  Moreover, for realistic physical systems inhomogeneity and
often uncontrollable interactions between nearby physical qubits pose
significant problems, rendering simple gate implementation schemes based
transversal application of single-qubit geometric pulse sequences
ineffective.  

Using an optimal control approach we can in principle overcome these
problems and find effective control pulse sequences to implement encoded
logic gates even for imperfect systems subject to limited control, e.g.,
if we are unable to selectively address individual qubits, or switch off
unwanted interactions between qubits.  Perhaps even more importantly,
the implementation of gates that are difficult to implement using
conventional techniques such as the $T$-gate for the five-qubit
stabilizer code, appears to present no greater challenge for optimal
control than the implementation of any other gate.  A potential drawback
at present is the complexity of the optimal control solutions.  However,
the optimal control solutions are not unique, and the solutions found
here should therefore be regarded more as a demonstration of principle,
i.e., that is is possible to find solutions to the optimal control
problem.  Simpler solutions are likely to exist at least for certain
problems, and future work exploring whether such solutions can be found
by considering different field parameterizations or imposing suitable
penalty terms that steer the algorithm away from complicated fields
towards simpler solutions (if they exist) could be fruitful.

\ack
SGS acknowledges funding from an EPSRC Advanced Research Fellowship and
Grant EP/D07195X/1, Hitachi and NSF Grant PHY05-51164, and would like to
thank Simon Devitt, Lorenza Viola, Pierre de Fouquiers and the participants
of the KITP Program on Quantum Control for valuable discussions.

\section*{References}

\bibliographystyle{prsty}
\bibliography{references}

\appendix
\section{Derivation of Eq.~(\ref{eq:DeltaU})}  
\label{appendix:A}

Letting for convenience
\begin{eqnarray*}
  H_0'(t) &=& H_0 + \sum_m u_m(t) H_m \\
  H_1'(t) &=& \sum_m \Delta u_m(t) H_m.
\end{eqnarray*}
we can rewrite the Schrodinger equation for $U_{\vec{u}}(t,t_0)$ and
$U_{\vec{u}+\vec{\Delta u}}(t,t_0)$ as
\begin{eqnarray*}
  \dot{U}_{\vec{u}}(t,t_0) 
  &=& -\frac{i}{\hbar}  H_0'(t) U_{\vec{u}}(t,t_0), \\
  \dot{U}_{\vec{u}+\vec{\Delta u}}(t,t_0) 
  &=& -\frac{i}{\hbar} [ H_0'(t) + H_1'(t)] U_{\vec{u}+\vec{\Delta u}}(t,t_0).
\end{eqnarray*}
Inserting an interaction picture decomposition
\begin{equation} 
 \label{eq:UI}
  U_{\vec{u}+\vec{\Delta u}}(t,t_0) =U_{\vec{u}}(t,t_0)U_I(t,t_0)
\end{equation} 
into the Schrodinger equation we obtain using the product rule
\begin{equation*}
  \dot{U}_I(t,t_0) = 
 -\frac{i}{\hbar} [U_{\vec{u}}(t,t_0)^\dag H_1'(t) U_{\vec{u}}(t,t_0)] U_I(t,t_0).
\end{equation*}
Rewriting this differential equation in integral form gives
\begin{equation*}
  U_I(t,t_0) = \ONE - \frac{i}{\hbar}
  \int_{t_0}^t [U_{\vec{u}}(\tau,t_0)^\dag H_1'(\tau)
  U_{\vec{u}}(\tau,t_0)] U_I(\tau,t_0)\, d\tau
\end{equation*}
Multiplying both sides by $U_{\vec{u}}(t,t_0)$ and noting that
$U_{\vec{u}}(t,t_0)$ does not depend on $\tau$ and thus can be
moved inside the integral now yields
\begin{eqnarray*}
\fl U_{\vec{u}+\vec{\Delta u}}(t,t_0)
 &=& U_{\vec{u}}(t,t_0) -\frac{i}{\hbar}\int_{t_0}^t 
     U_{\vec{u}}(t,t_0) U_{\vec{u}}(\tau,t_0)^\dag 
     H_1'(\tau) U_{\vec{u}}(\tau,t_0) U_I(\tau,t_0)\, d\tau \\
\fl &=& U_{\vec{u}}(t,t_0) -\frac{i}{\hbar}\int_{t_0}^t 
     U_{\vec{u}}(t,\tau) 
     H_1'(\tau) U_{\vec{u}+\vec{\Delta u}}(\tau,t_0) \, d\tau \\
\fl &=& U_{\vec{u}}(t,t_0) -\frac{i}{\hbar}\int_{t_0}^t 
     \sum_m \Delta u_m(t) U_{\vec{u}}(t,\tau) 
     H_m U_{\vec{u}+\vec{\Delta u}}(\tau,t_0) \, d\tau \\
\end{eqnarray*}
where we have used
$U_{\vec{u}}(t,t_0) U_{\vec{u}}(\tau,t_0)^\dag=U_{\vec{u}}(t,\tau)$ we
as well as Eq.~(\ref{eq:UI}) and the definition of $H_1'(t)$.

\section{Derivation of Eq.~(\ref{eq:DeltaF})}
\label{appendix:B}

If $\Delta u_m(t)=0$ except for $t\in I_k=[t_{k-1},t_k]$ then we have
\begin{eqnarray*} 
 \fl \Delta\F(t_F) 
     &=& \frac{2}{N\hbar}\sum_{m}\int_{t_{0}}^{t_F}\!\!\! 
     \Delta u_{m}(\tau)\Im
     \Tr[W^{\dag}U_{\vec{u}}(t_F,\tau)H_{m}U_{\vec{u+\Delta u}}(\tau,t_0)]\, d\tau\\
 \fl &=& \frac{2}{N\hbar}\sum_{m}\int_{t_{k-1}}^{t_{k}}\!\!\! 
     \Delta u_{m}(\tau)\Im
     \Tr[W^{\dag}U_{\vec{u}}(t_F,\tau)H_{m}U_{\vec{u+\Delta u}}(\tau,t_0)]\, d\tau
\end{eqnarray*}
Furthermore, we have $U_{\vec{u}}(t_F,\tau) =U_{\vec{u}}(t_F,t_k)
U_{\vec{u}}(t_k,\tau)$ as well as $U_{\vec{u+\Delta u}}(\tau,t_0)
=U_{\vec{u+\Delta u}}(\tau,t_{k-1}) U_{\vec{u+\Delta u}}(t_{k-1},t_0)$ and
because $U_{\vec{u}}(t_F,t_k)$ and $U_{\vec{u+\Delta u}}(t_{k-1},t_0)$ are
outside of the domain of integration we can re-write the expression for
$\Delta\F(t_F)$ as
\begin{equation*} \fl
\Delta\F(t_F) 
 = \frac{2}{N\hbar}\sum_{m}
     \Delta u_{m}(\tau)\Im
     \Tr\left[ W^{\dag}U_{\vec{u}}(t_F,t_k)
     \Delta U_m^{(k)} U_{\vec{u+\Delta u}}(t_k,t_0)\right] 
\end{equation*}
with $\Delta U_m^{(k)}$ as in (\ref{eq:DeltaUmk}).  Finally noting that
\begin{eqnarray*}
  U_{\vec{u}}(t_F,t_k) 
   &=& U_{\vec{u}}^{(K)} \cdots U_{\vec{u}}^{(k+1)} \\
  U_{\vec{u+\Delta u}}(t_k,t_0) 
   &=& U_{\vec{u}+\Delta\vec{u}}^{(k-1)} \ldots 
      U_{\vec{u}+\Delta\vec{u}}^{(1)}
\end{eqnarray*}
with $U_{\vec{u}}^{(k)}$ and $U_{\vec{u}+\Delta\vec{u}}^{(k)}$ as
defined in (\ref{eq:Uk}) gives
\begin{equation*}\fl
 \Delta\F(t_F) 
  = \frac{2}{N\hbar}\sum_{m} \Delta u_{mk} \Im \Tr \left[
     W^\dag U_{\vec{u}}^{(K)}  \cdots U_{\vec{u}}^{(k+1)} 
      \Delta U_m^{(k)} 
      U_{\vec{u}+\Delta\vec{u}}^{(k-1)} \ldots 
      U_{\vec{u}+\Delta\vec{u}}^{(1)}\right]
\end{equation*}

\end{document}